\documentclass[letterpaper,10pt,nofootinbib,aps,tightenlines,twocolumn]{revtex4}

\usepackage{amsmath,amsfonts,amssymb}
\usepackage{mathrsfs}
\usepackage{graphicx}
\usepackage[english]{babel} 
\usepackage{color}
\usepackage{simplewick}

\usepackage{appendix}

\usepackage{epstopdf}

\def\hatn{\mathbf{\widehat n}}

\newcommand{\beq}{\begin{equation}}
\newcommand{\eeq}{\end{equation}}
\newcommand{\bga}{\begin{gathered}}
\newcommand{\ega}{\end{gathered}}
\newcommand{\beqa}{\begin{eqnarray}}
\newcommand{\eeqa}{\end{eqnarray}}

\newcommand{\CAMB}{CAMB}

\begin{document}
\title{First CMB Constraints on Direction-Dependent Cosmological
Birefringence from WMAP-7}
\author{Vera Gluscevic$^1$, Duncan Hanson$^{1,2}$, Marc
     Kamionkowski$^{1, 3}$, and Christopher M. Hirata$^1$}
\affiliation{$^1$California Institute of Technology, Mail Code 350-17,
     Pasadena, CA 91125, USA\\
$^2$Jet Propulsion Laboratory, California Institute of Technology,
4800 Oak Grove Drive, Pasadena CA 91109, USA\\
$^3$Department of Physics and Astronomy, Johns Hopkins University, Baltimore, MD 21218, USA}
\date{\today}
\begin{abstract}
A Chern-Simons coupling of a new scalar field to
electromagnetism may give rise to cosmological birefringence,
a rotation of the linear polarization of electromagnetic waves as
they propagate over cosmological distances.  Prior work has
sought this rotation, assuming the rotation angle to be uniform
across the sky, by looking for the parity-violating TB and EB
correlations a uniform rotation produces in the CMB
temperature/polarization.  However, if the scalar field that
gives rise to cosmological birefringence has spatial
fluctuations, then the rotation angle may vary across the sky.  Here we search for
direction-dependent cosmological birefringence in the WMAP-7
data. We report the first CMB constraint on the rotation-angle
power spectrum $C_L^{\alpha\alpha}$ for multipoles between $L=0$
and $L=512$. We also obtain a $68\%$ confidence-level upper limit of
$\sqrt{C_2^{\alpha\alpha}/(4\pi)}\lesssim 1^{\circ}$ on the quadrupole
of a scale-invariant rotation-angle power spectrum.
\end{abstract}
\maketitle
\section{Introduction}
\label{sec:intro}
In this work, we use the cosmic microwave background (CMB)
temperature and polarization maps of the Wilkinson Microwave
Anisotropy Probe (WMAP) 7-year data
release \cite{Jarosik:2010iu} to search for direction-dependent cosmological
birefringence (CB). CB is a postulated rotation
of the linear polarization of photons that propagate through
cosmological distances \cite{Carroll:1989vb}. It
is present, for example, in models where a Nambu-Goldstone boson
plays the role of quintessence \cite{Carroll:1998zi}, but also
in models with new scalar degrees of freedom that have nothing
to do with quintessence
\cite{Li:2008tma,Pospelov:2008gg,Caldwell:2011pu,Finelli:2008jv}.
The rotation of the polarization is a consequence of the
coupling of a scalar field to the electromagnetic Chern-Simons
term, such that the rotation angle $\alpha$ is proportional to the total
change $\Delta\phi$ of the field $\phi$ along the photon's path.

Prior to this work, a rotation angle $\alpha$ that is uniform
across the sky had been sought in the CMB \cite{earlysearches},
where it would induce parity-violating TB and EB
temperature/polarization correlations \cite{Lue:1998mq}.  CB has
also been sought in quasar data
\cite{quasarsearches,Carroll:1989vb}.  The tightest constraint
currently comes from a combined analysis of the WMAP, Bicep
\cite{Chiang:2009xsa}, and QUAD experiments \cite{Wu:2008qb}; it
is $- {1.4^ \circ } < \alpha  < {0.9^ \circ }$ at the $95\%$
confidence level \cite{Komatsu:2010fb}.

There are, however, a number of reasons to expand the search and look for
a CB angle $\alpha(\hatn)$ that varies as a function of position
$\hatn$ on the sky.  To begin with, a
dynamical field $\phi$ that drives the rotation can have
fluctuations, in which case the rotation angle varies across the
sky
\cite{Li:2008tma,Pospelov:2008gg,Caldwell:2011pu}. Furthermore,
if $\phi$ is some massless scalar, not necessarily quintessence,
its background value does not necessarily evolve, and the
uniform component of the rotation angle may vanish.  The
only way to look for CB in this
scenario is through its direction dependence. Additionally, if
$\alpha(\hatn)$ is measured with high significance, the exact
shape of its power spectrum provides a window into the detailed
physics of the new cosmic scalar $\phi$.  Currently, the
strongest limit on a direction-dependent CB angle comes from AGN
\cite{Kamionkowski:2010ss}, which constrain the root-variance
of the rotation angle to be $\lesssim 3.7^\circ$.

In previous studies \cite{Kamionkowski:2008fp,
Gluscevic:2009mm}, a formalism was developed to search for
anisotropic CB rotation with the CMB.  The sensitivity of WMAP
data to this anisotropic rotation is expected to be competitive
with that from AGN
\cite{Kamionkowski:2008fp,Gluscevic:2009mm,Yadav:2009eb}.
However, the CMB also allows individual multipoles
$C_L^{\alpha\alpha}$ to be probed---the AGN data only constrain
the variance---and is sensitive to higher $L$ than AGN.  The CMB
also probes CB to a larger lookback time than AGN.

Here we apply the formalism developed earlier to the WMAP
7-year data.  Within experimental precision,
we report a non-detection of a direction-dependent cosmological
birefringence.  We obtain an upper limit on all the rotation-angle
power-spectrum multipoles $C_L^{\alpha\alpha}$ up to $L=512$.
This result implies a $68\%$ confidence-level upper limit on the quadrupole of a
scale-invariant power spectrum of
$\sqrt{C_2^{\alpha\alpha}/(4\pi)}\lesssim
1^{\circ}$,\footnote{Here, the power spectrum is defined in
the usual way, $C_L^{\alpha\alpha}\equiv \sum\limits_{M}
\alpha_{LM}\alpha_{LM}^*/(2L+1)$, where a spherical-harmonic
decomposition of the rotation field provides the rotation-angle
multipoles, $\alpha_{LM}\equiv\int Y_{LM}^*(\hatn) \alpha(\hatn)d\hatn$.}.
As a check, we also find a constraint on the uniform rotation
that agrees with the results of Ref.~\cite{Komatsu:2010fb}.

The rest of this paper is organized as follows. In
\S\ref{sec:physics}, we review the physical mechanism for
CB. In \S\ref{sec:formalism}, we revisit the full-sky
formalism to search for direction-dependent rotation, and discuss its implementation. WMAP data selection, our simulations, and the tests of the analysis method are described in
\S\ref{sec:data_simulations}. Results are reported in
\S\ref{sec:constraints}, and we conclude in
\S\ref{sec:discussion}. Appendix \ref{ax:small_alpha} contains a detailed explanation of the procedure we used to obtain an upper limit of the root-mean-squared rotation angle from the measurement of the $TE$ correlation in the data; Appendix \ref{ax:kernels} contains a discussion of the geometrical properties of the rotation-angle estimator; Appendix \ref{ax:Lfsky} details the calculation of the $L$-dependence of the fractional correction for a scale-invariant power spectrum recovered from cut-sky maps; and Appendix \ref{ax:masks} displays the analysis masks we used in this work.
\section{Physical origin of cosmological birefringence}
\label{sec:physics}
Theories with a weakly broken global $U(1)$ symmetry provide a natural mechanism for producing a shallow potential for the pseudo-Nambu-Goldstone-boson field $\phi$. From a cosmological perspective, a PNGB field with this property is a natural candidate for quintessence, since it can drive epochs of accelerated expansion \cite{Carroll:1998zi}. In addition, many other extensions of the Standard Model of
particle physics and $\Lambda$CDM cosmology abound in scalar fields descending
from theories with shift symmetry and resembling the PNGB. Such
fields generically couple to photons through the Chern-Simons
term $F^{\mu\nu}\widetilde{F}_{\mu\nu}$ of electromagnetism,
while the underlying shift symmetry suppresses all other
leading-order couplings to Standard-Model particles
\cite{Carroll:1998zi}. This way, the existence of a new degree
of freedom $\phi$ could evade detection in colliders and other
lab experiments, but could still be manifest in cosmology
through CB. We now review in more detail the physical mechanism
that gives rise to CB.
 
The Chern-Simons--modified electromagnetic Lagrangian reads
\begin{eqnarray}
\label{eq:FFdual}
     \mathscr{L} &=&-\frac{1}{4}F^{\mu\nu}F_{\mu\nu}- \frac{\beta}{2M}\phi F^{\mu\nu}\widetilde{F}_{\mu\nu},
\end{eqnarray}
where $F^{\mu\nu}$ is the electromagnetic field-strength tensor, $\widetilde F^{\mu\nu}$ is its dual, $\beta$ is a coupling constant, and the mass $M$ is a vacuum expectation value of the spontaneously broken symmetry. The dispersion relation following from this modified Lagrangian has different solutions for the left- and right-handed photon polarizations, the net effect being the rotation of the linearly polarized electromagnetic wave that propagates through the vacuum with the evolving field $\phi$. The direction of polarization is rotated by an amount
\beq
\alpha=\frac{\beta}{2M}\Delta \phi,
\label{eqn:deltaalpha}
\eeq
that depends on the total change $\Delta\phi$ along the photon's
path. Since $M$ can be arbitrarly large, perhaps on the order of
the Planck mass, the accumulated change in $\phi$ must also be large in order for this angle to be measurable. This motivates
the use of cosmological probes in search for CB. There are models in which
$\Delta\phi$ is uniform across the sky (giving rise only to a
uniform rotation angle), as well as models in which it has
anisotropies~\cite{Li:2008tma,Pospelov:2008gg,Caldwell:2011pu}.
In this work, we do not focus on any particular
physical model for CB, but rather derive a model-independent
constraint on the rotation-angle power spectrum $C_L^{\alpha\alpha}$. 
\section{Full-sky formalism and its implementation}
\label{sec:formalism}
In this Section we review the full-sky-estimator formalism
of Ref.~\cite{Gluscevic:2009mm} for measuring direction-dependent CB. In order to apply this formalism to the cut sky (after masking out the Galaxy), all measured power spectra need to be corrected by a factor of $\sim1/f_\text{sky}$.\footnote{When multipole coefficients are calculated from a map where a fraction $1 - f_\text{sky}$ of the pixels is masked (i.e. signal set to zero), the usual full-sky expression for their variance (i.e. the power spectrum; see Ref.~\cite{Kamionkowski:1996ks} or Eq.~\eqref{eq:caa_biased} where $f_\text{sky}=1$) is underestimated by a factor of $f_\text{sky}$, because the variance corresponding to the masked pixels is effectively zero.} Unless otherwise noted, $f_\text{sky}$ is calculated as the fraction of the pixels that the mask admits; we include this factor when appropriate in the following derivation. We
rewrite all the relevant formulas in a position-space form which is numerically
efficient for the analysis of data. 

In the presence of birefringence, the polarization field acquires a phase factor,
\begin{equation}
    p(\hatn)\equiv [Q+iU](\hatn) = \widetilde p(\hatn) e^{2i\alpha(\hatn)},
\label{eq:qiu_observed}
\end{equation}
where $Q$ and $U$ are the
Stokes parameters for linear polarization, and tilde denotes the
polarization in the absence of birefringence, which we
refer to as the ``primary polarization''. 
To obtain an estimate of the phase factor $e^{2i\alpha(\hatn)}$ from the
polarization field in Eq.~\eqref{eq:qiu_observed}, we require a
tracer of $\widetilde p(\hatn)$. The primary
polarization is generated by Thomson scattering of the local
temperature quadrupole, so the observed temperature field $T(\hatn)$ may be used for this purpose. Due to projection
effects, the local temperature quadrupoles at last scattering
appear on the sky as a curvature of the temperature field. The estimator for the rotation angle then involves projecting the temperature field into a map as a spin-2 quantity (which evaluates the curvature), and looking for correlation with the polarization field which varies as a function of the position on the sky. We review a rigorous derivation of the estimator in the following subsections.  
\subsection{Rotation-induced B modes}
\label{sec:rotation_induced_bmodes}
On the full sky, the polarization field can be decomposed in terms of spin-2 spherical harmonics ${}_2Y_{lm}^{}(\hatn)$ as
\beq
p(\hatn)=- \sum\limits_{lm} {({E_{lm}} + i{B_{lm}})} {}_2Y_{lm}^{}(\hatn),
\eeq
where $E$ and $B$ modes represent polarization patterns of opposite parity \cite{Kamionkowski:1996ks,Zaldarriaga:1996xe}.

The primary $E$-mode polarization signal $\widetilde E_{lm}$ (sourced by the
dominant scalar perturbations) is detected with high
significance in WMAP-7 data \cite{Komatsu:2010fb}, although
primary $B$ modes (sourced by the subdominant tensor
perturbations) have only been constrained with upper limits.
For this reason, most of the constraining power for CB in WMAP
comes from the search for a CB-induced rotation of the primary
$E$ mode into an observed $B$ mode. The induced $B$ mode is given as
\cite{Kamionkowski:2008fp,Gluscevic:2009mm}
\begin{multline}
B_{lm}  =\frac{i}{2} \int {} d\hatn [\widetilde p(\hatn)e^{2i\alpha(\hatn)}{}_2Y_{lm}^*(\hatn) \\
- \widetilde p(\hatn)^*e^{-2i\alpha(\hatn)}{}_{ - 2}Y_{lm}^*(\hatn)].
\end{multline}
This $B$ mode is correlated with the primary $E$ mode (from which it originated), and through it also with the temperature anisotropies. The presence of rotation therefore gives rise to anomalous $EB$ and $TB$ correlations, and both these power spectra can be used to search for CB.
It is, however, worth keeping in mind that individual multipoles of the $E$-mode polarization signal are still noise dominated, whereas the temperature is measured at $S/N>1$, for a large number of multipoles, in every frequency band of the WMAP 7-year data.
Therefore, at WMAP noise levels the temperature field makes a better tracer of the primary $E$-mode than the observed $E$ mode itself. For this reason, on most angular scales, the search for a $TB$ correlation, which we focus on in this work, provides the best constraint on CB \cite{Gluscevic:2009mm}.
Assuming the primary polarization is a pure E mode at the surface of last scatter, the CB-induced $TB$ correlation reads \cite{Kamionkowski:2008fp,Gluscevic:2009mm}
\begin{multline}
\left\langle B_{lm}T_{l'm'}^* \right\rangle  = \int d\hatn \widetilde C_{l'}^{TE}\\
\times\Bigg[\frac{1}{2}\rm{sin}(2\alpha)[{}_2Y_{l'm'}{}_2Y_{lm}^* + {}_{ - 2}Y_{l'm'}{}_{ - 2}Y_{lm}^*]\\
- \frac{i}{2}\rm{cos}(2\alpha)[{}_2Y_{l'm'}{}_2Y_{lm}^* - {}_{ - 2}Y_{l'm'}{}_{ - 2}Y_{lm}^*]\Bigg],
\label{eq:TB_whole}
\end{multline}
where we suppress the $\hatn$ dependence for clarity. The two contributions to the correlation, sin and cos, have opposite parities, where only terms that satisfy $l\!+\!l'\!+\!L\!=$even, and $l\!+\!l'\!+\!L\!=$odd, respectively, contribute to the sum. Power spectrum $\widetilde C_{l'}^{TE}$ is the correlation between the temperature and the primary $E$
mode, which may be calculated with \CAMB~\cite{Lewis:1999bs}.

So far, we have not assumed anything about the magnitude of the rotation per pixel in CMB maps. Observations of quasars suggest an upper bound on the root-mean-squared (RMS) of $\alpha(\hatn)$ of just a few degrees \cite{Kamionkowski:2010ss}, while the measurement of the $TE$ correlation from WMAP-7 data implies a somewhat weaker constraint: $\left<\alpha(\hatn)^2\right> < 9.5^\circ$ (see Appendix \ref{ax:small_alpha} for details). Motivated by these results, in the rest of this paper we adopt a small--rotation-angle limit. The numerical results we present in \S\ref{sec:claa_constraints} do not depend on the validity of this assumption, but their interpretation as an upper limit of the rotation-angle autocorrelation $C_L^{\alpha\alpha}$ does; this subtlety is discussed in more detail in \S\ref{sec:data_simulations}, \S\ref{sec:constraints}, and Appendix \ref{ax:small_alpha}. 

In the limit of small rotation angle, only the sine term contributes to the observed $TB$ which then reads
\begin{multline}
\left\langle B_{lm}T_{l'm'}^* \right\rangle\approx \int d\hatn \widetilde C_{l'}^{TE}\alpha(\hatn)\\
\times[{}_2Y_{l'm'}(\hatn){}_2Y_{lm}^*(\hatn) + {}_{ - 2}Y_{l'm'}(\hatn){}_{ - 2}Y_{lm}^*(\hatn)],
\label{eq:TB}
\end{multline}
A $TB$ correlation generated by weak gravitational lensing of the CMB is of opposite parity, with $l\!+\!l'\!+\!L\!=$odd, and does not represent a source of bias for measuring a small-rotation signal. 
In addition, the effect of lensing is not internally observable at WMAP noise levels, even with an optimal estimator \cite{Smith:2007rg}. We therefore do not consider lensing further.
\subsection{Minimum-variance quadratic estimator: $\widehat \alpha_{LM}$}
\label{sec:mve}
From Eq.~\eqref{eq:TB}, it is evident that scale-dependent birefringence induces
correlations between temperature and polarization
modes at different wavenumbers $l$, $l'$; i.e., it produces a statistically anisotropic imprint on
the covariance matrix of the observed CMB. Each $ll'$ pair
measured in the maps may therefore be used as an estimate of the
rotation-angle multipole $\alpha_{LM}$, provided that it
satisfies the usual triangle inequalities, $L\leq l+l'$ and
$L\geq |l-l'|$, as well as the parity condition
$l\!+\!l'\!+\!L\!=$even. The prescription for combining all $ll'$
estimates in order to produce a minimum-variance quadratic
estimator is explained in detail in
Ref.~\cite{Gluscevic:2009mm}. Here, we only present the final
expressions for the $TB$ estimator,
\begin{multline}
\widehat \alpha _{LM}^{} = N_L \int {} d\hatn {Y_{LM}(\hatn)}\\
\times\Bigg[ \sum\limits_{lm} {\bar B_{lm}^*} {}_2Y_{lm}(\hatn) \sum\limits_{l'm'} \widetilde{C}_{l'}^{TE}\bar T_{l'm'} {}_2Y_{l'm'}^{*}(\hatn)\\
+ \left( \text{complex conjugate}\right)\Bigg] ,
\label{eq:estimator}
\end{multline}
where $N_L$ is an $L$-dependent normalization and the barred quantities represent inverse-variance filtered multipoles. For full-sky coverage and homogenous noise in pixel space, the expressions for these quantities read
\beq
\bga
\bar B_{lm} \equiv \frac{{B_{lm}^{}}}{{C_l^{BB}}}, \quad \bar T_{l'm'}^{} \equiv \frac{{T_{l'm'}^{}}}{{C_{l'}^{TT}}}.
\ega
\label{eq:IVF_bars}
\eeq
The unbarred $B$ and $T$ are the observed temperature and polarization multipoles corrected for the combined instrumental beam and pixelization transfer function $W_l$, and the $TT$ and $BB$ power spectra are analytic estimates of the total signal +  noise power spectrum in a given frequency band,
\beq
\bga
C^{TT}_l \equiv \widetilde C^{TT}_l + C_l^{TT,\text{ noise}}/W^2_l,\\
C^{BB}_l \equiv \widetilde C^{BB}_l + C_l^{BB,\text{ noise}}/W^2_l.
\ega
\label{eq:ivf}
\eeq
In the idealized case of full-sky coverage and homogeneous instrumental noise, the estimator normalization $N_L$ is calculable analytically and is equal to the inverse of the estimator variance,
\beq
\bga
N_L = \left( \sum_{ll'}\frac{(2l+1)(2l'+1)}{4\pi}\frac{(\widetilde C_{l'}^{TE})^2}{C_l^{BB}C_{l'}^{TT}}(H_{ll'}^L)^2\right)^{-1},
\ega
\label{eq:NL}
\eeq
where
\beq
\bga
{H_{ll'}^L} = \left( {\begin{array}{*{20}{c}}
   l  \\
   { - 2}  \\
\end{array}\begin{array}{*{20}{c}}
   L  \\
   0  \\
\end{array}\begin{array}{*{20}{c}}
   {l'}  \\
   2  \\
\end{array}} \right) + \left( {\begin{array}{*{20}{c}}
   l  \\
   2  \\
\end{array}\begin{array}{*{20}{c}}
   L  \\
   0  \\
\end{array}\begin{array}{*{20}{c}}
   {l'}  \\
   { - 2}  \\
\end{array}} \right).
\ega
\label{eq:w3j_20m2}
\eeq
The objects in parentheses are Wigner-3j symbols.

In the non-idealized case of real data, the simple inverse-variance filters (IVFs) presented above are suboptimal, in the sense that the associated estimator variance is not truly minimized. To obtain a true minimum-variance estimate, computationally more involved filters are required \cite{Smith:2007rg}. In practice, however, we find that the full-sky expressions for the estimator in Eq.~\eqref{eq:estimator} provide a very good approximation to its behavior on the cut sky. Namely, the analytic expression for its variance, given by Eq.~\eqref{eq:NL}, is consistent with the full variance recovered from a suite of Monte Carlo simulations (described in detail in \S\ref{sec:simulations}) when the simple IVFs of Eq.~\eqref{eq:ivf} are used in the presence of sky cuts; the appropriate correction for the fraction of the sky admitted by the analysis masks must be included in this case. This result motivates us to continue using the simple IVFs and the corresponding analytic expressions for the estimator normalization. 

The insensitivity to the presence of the galaxy masks that we observe here can be interpreted as a consequence of the following properties. First, the estimator of Eq.~\eqref{eq:estimator} is a product of inverse-variance filtered $T$ and $B$ maps in real space, which are local functions of the data. The IVFs are local in pixel space (they resemble Gaussians with a width of a few arcmins,
corresponding to the resolution in a given frequency band), and
so the mask boundaries remain localized after
filtering. Additionally, the estimator is an even function of the
temperature map (see Eq.~\eqref{eq:estimator}---it contains a
second derivative of the temperature field performed by
${}_2Y_{l'm'}$), and so it is relatively insensitive to the
discontinuities introduced by the analysis mask. These properties put the rotation estimator $\widehat\alpha_{LM}$ in sharp contrast with the estimators for the gravitational-lensing potential, where the dependence on the gradient of the temperature field renders the lensing reconstruction very sensitive to sky cuts \cite{Hirata:2004rp}. 
\subsection{Power-spectrum estimator: $\widehat C_L^{\alpha\alpha}$}
\label{sec:caa_estimator}
Once the rotation-angle multipoles are measured, their autocorrelation can be estimated as
\beq 
C_L^{\widehat\alpha \widehat\alpha} \equiv \frac{1}{f_\text{sky}(2L + 1)}\sum\limits_M {{{\widehat \alpha }_{LM}}\widehat \alpha _{LM}^*}.
\label{eq:caa_biased}
\eeq
This represents a sum over the $\left<TBTB\right>$ trispectrum, where ``T'' and ``B'' denote the temperature and B-mode multipole moments. This estimator for $C_L^{\alpha\alpha}$ is non-zero even in the absence of CB-induced rotation, due to the presence of Wick contractions from the primary CMB and the instrumental noise (discussed in \S\ref{sec:mve}). They produce the ``noise bias" $C_L^{\alpha\alpha,\ \text{noise}}$ and must be subtracted from the measurement of  $C_L^{\widehat \alpha\widehat \alpha}$, in order to recover an estimate of the CB-induced signal $C_L^{\alpha\alpha}$,
\beq
\widehat{C}_L^{\alpha\alpha}= C_L^{\widehat \alpha \widehat \alpha} - C_L^{\alpha\alpha,\ \text{noise}}.
\label{eqn:claa_debiasing}
\eeq
For Gaussian CMB+noise fluctuations, the noise bias can be identified with the three disconnected Wick contractions of the trispectrum which $C_L^{\widehat\alpha\widehat\alpha}$ probes:
\beq
\begin{array}{c}
(a) \quad
\contraction{}{T}{}{B}
\contraction{TB}{T}{}{B}
TBTB \\
(b)\quad
\contraction{}{T}{B}{T}
\contraction[2ex]{T}{B}{T}{B}
TBTB \\
(c)\quad
\contraction{}{T}{BT}{B}
\contraction[2ex]{T}{B}{}{T}
TBTB.
\end{array}
\label{eqn:contractions}
\eeq
where we neglect contraction (a), which only couples to the $L=0$ mode of $\widehat{\alpha}_{LM}$, and also contraction (c), as it is negligible\footnote{In our simulations, we verify that this term has indeed a negligable numerical contribution.}. In the absence of statistical anisotropy (i.e. for full-sky coverage and homogeneous instrumental noise), the contraction (b) between two real fields with multipoles $lm,\ l'm'$ carries a set of delta functions $\delta_{ll'} \delta_{mm'}$, and the realization-dependent noise bias may be written explicitly in terms of the observed power spectra. If $C_L^{\widehat\alpha\widehat\alpha}$ is evaluated by cross-correlating the $f_1$, $f_2$, $f_3$, and $f_4$ frequency-band maps, the analytic expression for this ``isotropic bias'' follows from Eq.~\eqref{eq:estimator},
\beq
\bga
C_L^{\alpha^{f_1f_2}\alpha^{f_3f_4}\text{,noise,iso}} \equiv {\left\langle {\widehat \alpha _{LM}^{}\widehat \alpha _{LM}^{*}} \right\rangle }_{\text{Gauss,iso}} = \\
\sum_{ll'}\frac{(2l+1)(2l'+1)}{4\pi}(H_{ll'}^L\tilde C_{l'}^{TE})^2\\
\times\frac{C_{l'}^{TT,f_1f_3\text{maps}}C_{l}^{BB,f_2f_4\text{maps}}}{(C_{l'}^{TT,f_1f_1}C_{l}^{BB,f_2f_2}C_{l'}^{TT,f_3f_3}C_{l}^{BB,f_4f_4})},
\ega
\label{eq:iso_bias}
\eeq
where the power spectra in the denominator of Eq.~\eqref{eq:iso_bias} are the simple analytic IVFs. The $C_{l'}^{TT,f_1f_3\text{maps}}$ and $C_l^{BB,f_2f_4\text{maps}}$ are measured by cross-correlating data maps in the frequency bands $f_1$ and $f_3$, or $f_2$ and $f_4$ respectively, and corrected by the factor of $1/f_\text{sky}^\text{T}$ and $1/f_\text{sky}^\text{P}$, corresponding to the temperature and polarization analysis mask, respectively. Most of the power in temperature comes from CMB fluctuations, and the $B$-mode power is mostly noise if $f_2=f_4$, and negligible otherwise. Therefore, since the instrumental noise is independent for different frequency bands, the largest contribution to the noise bias can be eliminated by cross-correlating estimates of $\widehat\alpha_{LM}$ obtained from two different bands. 

In reality, we work with a masked sky which has been observed with inhomogeneous noise levels, and Eq.~\eqref{eq:iso_bias} does not provide a perfect description of the noise bias, although it is an excellent first approximation. This leads us to adopt a two-stage debiasing procedure in which we subtract both the isotropic bias of Eq.~\eqref{eq:iso_bias}, and an addiitonal Monte-Carlo--based correction, in order to correct for the effects of sky cuts and inhomogeneity of the instrumental noise. The total noise bias $C_L^{\alpha\alpha,\ {\rm noise}}$ is the sum of the two contributions,
\beq
C_L^{\alpha \alpha,\ {\rm noise} } \equiv C_L^{\alpha \alpha,\ {\rm noise},\ {\rm iso}} + C_L^{\alpha\alpha,\ {\rm noise},\ {\rm MC}}.
\eeq
We estimate $C_L^{\alpha\alpha,\ {\rm noise},\ {\rm MC}}$ from a set of WMAP realizations generated with no birefringence signal (described in \S\ref{sec:simulations}), analyzed in the same way as the data itself. For each realization, we calculate the appropriate $C_L^{\alpha \alpha,\ {\rm noise},\ {\rm iso}}$ and averaging over many realizations obtain $C_L^{\alpha\alpha,\ {\rm noise},\ {\rm MC}}$ as
\beq 
C_L^{\alpha\alpha,\ {\rm noise},\ {\rm MC}} \equiv \langle C_L^{\widehat\alpha \widehat\alpha} - C_L^{\alpha \alpha,\ {\rm noise},\ {\rm iso.}} \rangle_{{\rm sims}}.
\eeq
This two-stage procedure reduces the sensitivity of our estimator to uncertainties in the CMB and instrumental-noise model, as compared to the case where the entire bias is recovered from Monte Carlo analysis. With the two-stage procedure, the largest (isotropic) contribution to the bias is evaluated directly from the power spectra of the observed maps, and is specific to the CMB realization at hand; subtracting it from the bispectrum naturally takes care of any noise (bias) contribution that might arise from the uncertainty in the background cosmology or in the noise description used to generate Monte Carlo simulations.

As we show in \S\ref{sec:constraints}, we find consistency of results obtained with either (i) the calculation of the trispectrum as a four-point autocorrelation of the maps in the same band,  (ii) the calculation of the trispectrum from cross-band correlations, which have an almost negligible noise bias.
\section{Data and simulations}
\label{sec:data_simulations}
\subsection{Data}
\label{sec:data}
Our main results are based on the full-resolution (corresponding to HEALPix resolution of $N_\text{side}=512$) co-added seven-year sky maps that contain foreground-reduced measurements of the Stokes $I$, $Q$, and $U$ parameters in three frequency bands: Q (41 GHz), V (61 GHz), and W (94 GHz), available at the LAMBDA website \cite{lambda}. A summary of the instrumental parameters most relevant to this analysis is provided in Table \ref{tb:bands}.
\begin{table}[htbp]
\begin{center}
    \begin{tabular}{ | c | c | c | c | p{10cm} |}
    \hline
    Band & FWHM & $\Delta_T$ [$ \mu$K$\text{arcmin}$]& $\Delta_P$ [$\mu$K$\text{arcmin}$] \\ \hline
    Q (41 GHz) & 34' & 316 & 544 \\ \hline
    V (61 GHz) & 24' & 387 & 589 \\ \hline
    W (94 GHz) & 22' & 467 & 693 \\ \hline
    \end{tabular}
\caption{Relevant instrumental parameters: beam full-width half
maximum (FWHM) and approximate map noise for temperature and
polarization for the three frequency bands we use in the
analysis \cite{lambda}.}\label{tb:bands}
\end{center}
\end{table}
We apply the seven-year temperature $KQ85y7$ mask with $78.27\%$ of the sky admitted, and a polarization $P06$ mask with $73.28\%$ of the sky admitted. These masks are constructed to remove diffuse emission based on
the data in K and Q bands, and on a model of thermal dust
emission.  Point sources are masked based on a combination of external catalog data and WMAP-detected sources. (For more information about the exclusion masks, see Appendix \ref{ax:masks} and Ref.~\cite{Gold:2010fm}.) 
\subsection{Simulations}
\label{sec:simulations}
We produce a suite of simulated WMAP observations, both to test the normalization of our $\alpha_{LM}$ estimates as well as to estimate their variance for the subtraction of $C_L^{\alpha\alpha,\ {\rm noise}}$ in Eq.~\eqref{eqn:claa_debiasing}. We produce simple simulations of the WMAP data with the following procedure:
\begin{enumerate}
\item Generate CMB-sky temperature and polarization realizations for the best-fit \mbox{``LCDM+SZ+ALL''} WMAP-7 cosmology of Ref.~\cite{lambda}.
\item Convolve the simulated CMB skies with a symmetric experimental beam. For the WMAP band maps we use an effective beam calculated as the average beam transfer function for all differencing assemblies at the given frequency.
\item Add simulated noise realizations based on the published
$I, Q, U$ covariance matrices within each pixel. We do not make
any attempt to generate noise with pixel-to-pixel noise
correlations, although we do exclude multipoles with $l<100$
from our analysis, as this is where most of this correlated noise resides. In \S\ref{sec:claa_constraints} we demonstrate the consistency of $\widehat C_L^{\alpha\alpha}$ estimates constructed from auto- and cross-correlations of maps with independent noise realizations, and so are justified in neglecting correlated noise in our analysis.
\end{enumerate}
We do not include Galactic foreground residuals or unresolved point sources in our simulations, but we address their possible impact on our results in \S\ref{sec:systematics}. 
\subsection{Test runs}
\label{sec:tests}
In order to demonstrate the recovery of the CB signal using the minimum-variance estimator formalism and the de-biasing method discussed in the previous sections, we generate a suite of simulations that include a CB signal, i.e. where the polarization maps are rotated by realizations of a scale-invariant power spectrum of $\alpha$,
\beq
C_L^{\alpha\alpha} = AC_L^{\alpha\alpha\text{,fiducial}} \equiv A\frac{131\text{deg}^2}{L(L+1)},
\label{eq:caa_SI}
\eeq
where we choose the amplitude of this fiducial model so that it
gives a $S/N$ ratio of order $1$ at low $L$ for WMAP V band, and an RMS rotation-angle on the sky of $10^\circ$, satisfying the small-angle approximation\footnote{It is important that the model satisfies the small-angle approximation, as our calculation of the bias from Monte Carlo analysis is based on the null-assumption. In regime where this approximation is not satisfied, higher-order corrections will be necessary to recover the rotation-anlge power spectrum from the measured $<TBTB>$ trispectrum.}. We apply analysis masks to each simulated map, and then analyze the map cross-correlations (as described in previous sections), recovering $\widehat\alpha_{LM}$ multipoles; we then compute the power spectrum using Eq.~\ref{eq:caa_biased}. Due to the interaction of the power distribution at different scales in the map with the geometry of the analysis mask, the $f_\text{sky}$ factor is in principle a function of the multipole moment $L$, which typically starts smaller than the average\footnote{The ``average" here is the usual $f_\text{sky}$ fraction associated with a mask, equal to the fraction of the pixels that the mask admits.} value at low $L$'s, and converges towards the average value at high $L$'s. Since most of the signal for this model (which we come back to in \S\ref{sec:SI_constraints}) comes from low $L$'s, we evaluate the exact $L$ dependence, and substitute the $f_\text{sky}(L)$ function in Eq.~\ref{eq:caa_biased} (for more details on $f_\text{sky}(L)$, see Appendix \ref{ax:Lfsky}). In Fig.~\ref{fig:tests}, show the results of this test, comparing the input $C_L^{\alpha\alpha}$ power spectrum to the mean of the reconstructed power from a large number of simulations, and demonstrate a successful recovery of the signal. In the following Section, we apply the same signal-reconstruction method to WMAP-7 data.

To conclude this section, we note one subtlety necessary for the correct interpretation of the results of our analysis. The expression for the estimator of Eq.~\eqref{eq:estimator} only recovers the rotation-angle multipole in the small-angle regime. In the general case of arbitrarily large rotation, Eq.~\eqref{eq:estimator} provides an exact estimate of the multipoles of another ``observable'' quantity: $\frac{1}{2}\rm{sin}[2\alpha(\hatn)]$. Strictly speaking, our de-biasing procedure also relies on the small-angle approximation, since $C_L^{\alpha\alpha\text{, noise, MC}}$ is calculated from a suite of null simulations. It is therefore necessary to inquire which regime corresponds to a particular model of rotation before interpreting our results as a constraint on such a model. However, the fiducial model we use as an example here (and which we come back to in \S\ref{sec:SI_constraints}) satisfies this assumption (producing an RMS rotation of $\sim 10^\circ$). In this particual case, the difference between the two power spectra of $\alpha$ and of $\frac{1}{2}\text{sin}[2\alpha]$ is mainly contained in the $15\%$ difference in their amplitudes. It thus is possible to recover the rotation-angle power spectrum by simply rescaling the measured power---the fact we use in \S\ref{sec:SI_constraints} to constrain this model from WMAP data. 
\begin{figure}[htbp]
\includegraphics[height=7cm,keepaspectratio=true]{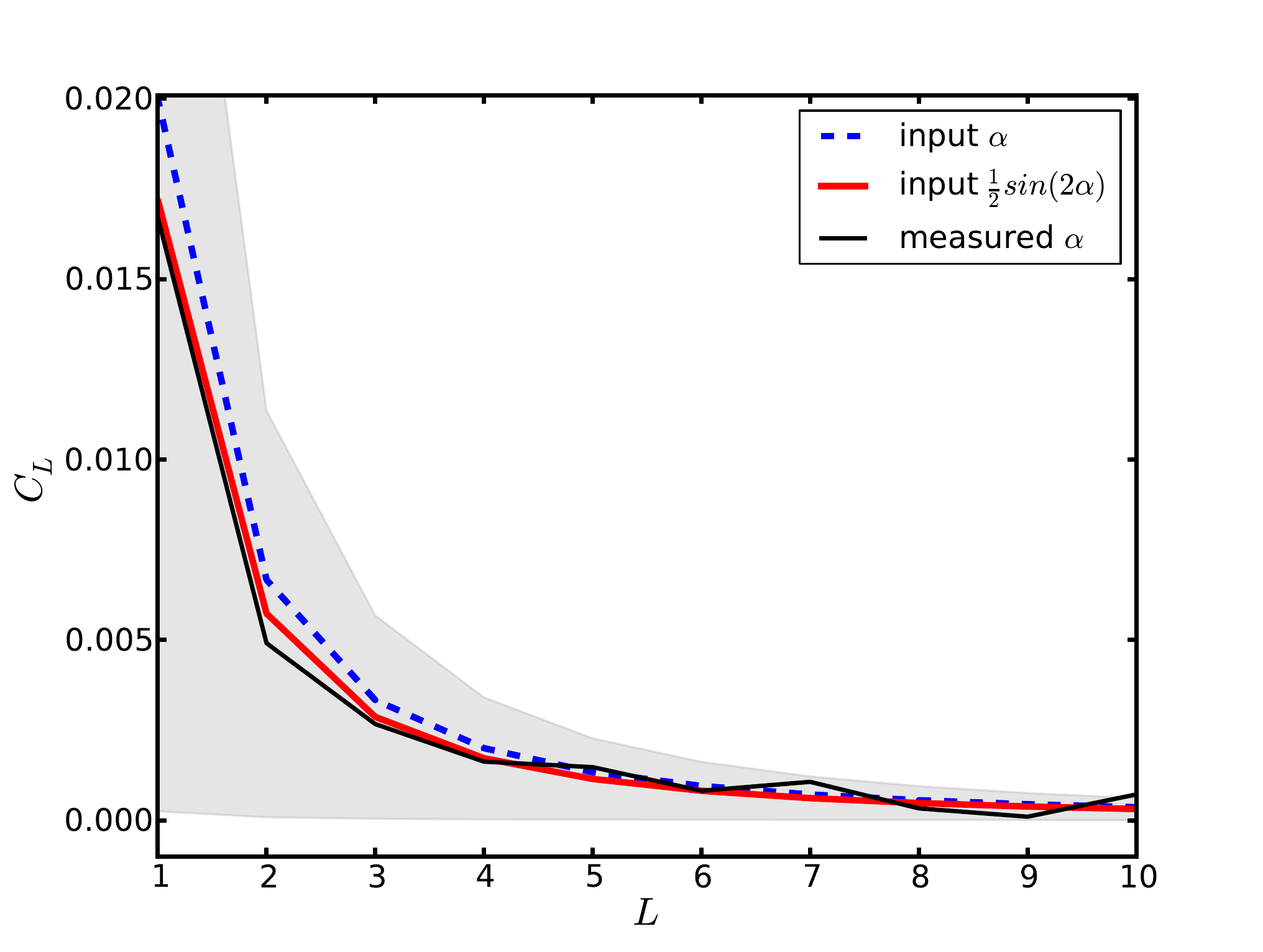}
\caption{The recovery of the CB signal with the analysis method presented in \S\ref{sec:formalism} is demonstrated using a suite of simulations that include realization of a rotated sky. Blue dashed line is the input signal power-spectrum of $C_L^{\alpha\alpha}$, red thick line represents the power spectrum of $\frac{1}{2}\rm{sin}[2\alpha(\hatn)]$, and the thin black line is the mean recovered power from the simulations; the gray region is a $1\sigma$--confidence-level interval calculated from the same suite of simulations. 
\label{fig:tests}}
\end{figure}
\section{Results}
\label{sec:constraints}
\subsection{Model-independent constraints}
\label{sec:claa_constraints}
Before continuing, let us first clarify our notation. The rotation-angle power spectra are marked with four frequency bands as [$f_1f_2$][$f_3f_4$]. This means that the two estimates of $\alpha_{LM}$ needed to evaluate the power spectrum are obtained by cross-correlating band $f_1$ with $f_2$, and $f_3$ with $f_4$, respectively. Here, the temperature multipoles are measured from $f_1$ and $f_3$, and the B modes are obtained from the maps in $f_2$ and $f_4$ bands. We measure five different band cross-correlations: [VV][VV], [QV][QV], [QQ][VV], [WV][WV], and [WW][VV], but since the results for all of them are qualitatively the same, here we only show plots for a characteristic subset. 

\begin{figure}[htbp]
\includegraphics[height=7cm,keepaspectratio=true]{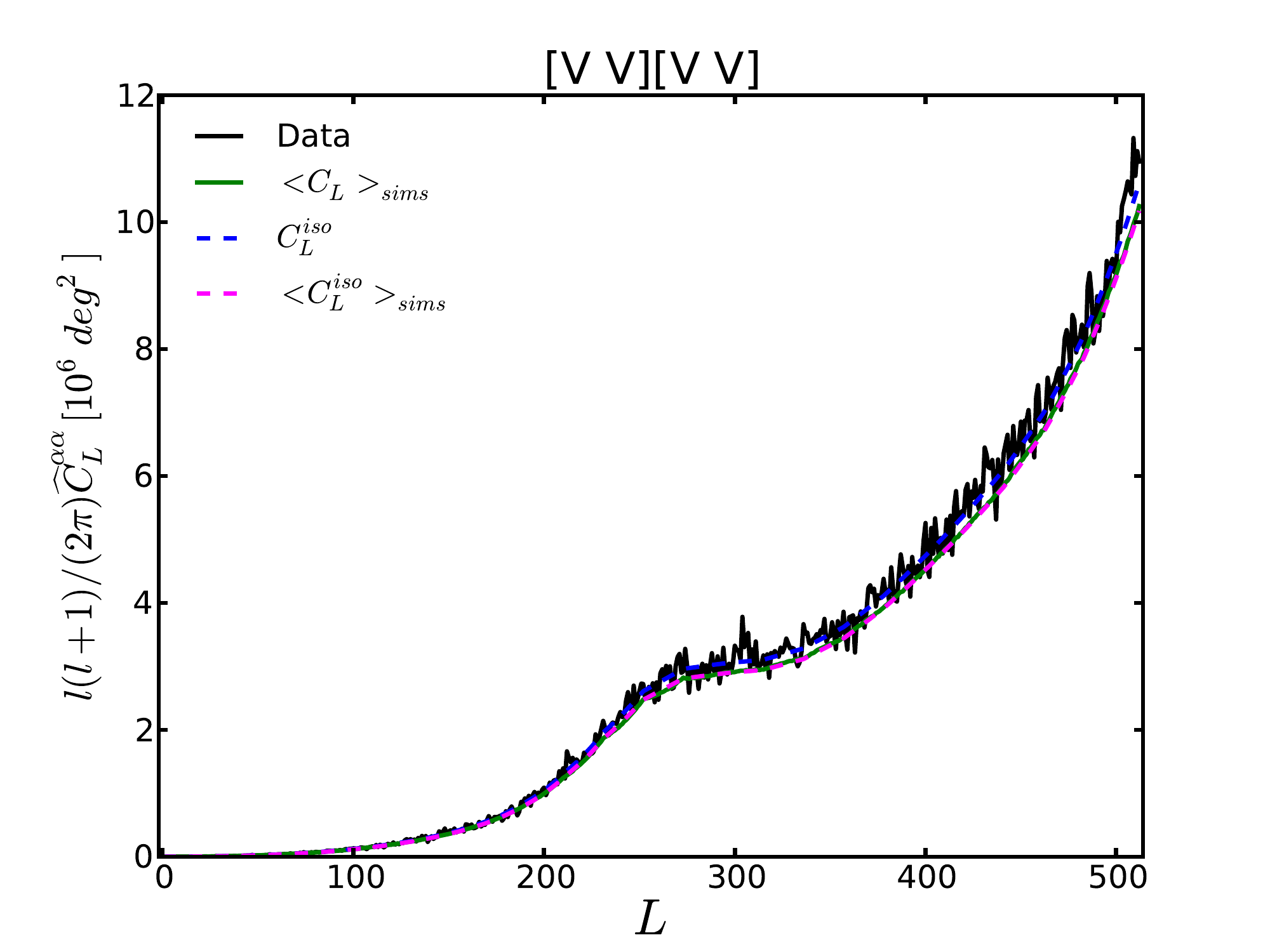}
\includegraphics[height=7cm,keepaspectratio=true]{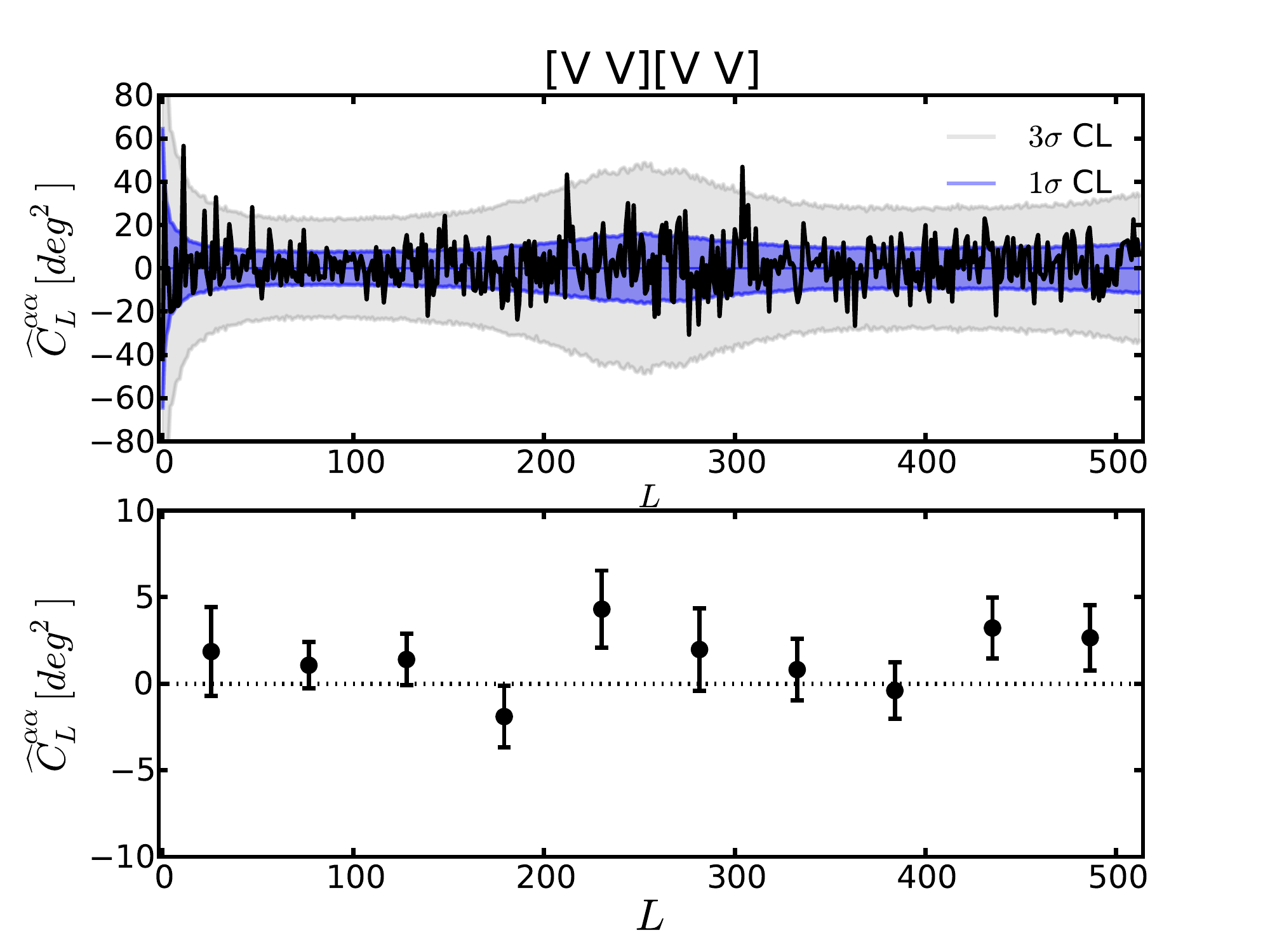}
\caption{Top panel: Measurement of the rotation-angle power spectrum from V band, shown before debiasing, along with the components of the noise bias: Monte-Carlo measurement of the null-hypothesis mean $\langle \widehat C_L^{\alpha\alpha} \rangle$ (solid green), isotropic noise bias (blue dashed), and the mean isotropic bias (magenta dashed). Middle panel: The same power spectrum after debiasing, with $1\sigma$ and $3\sigma$ confidence interval. Bottom panel: binned version of the middle-panel power spectrum. The results are consistent with zero within $3\sigma$.
\label{fig:claa_VVVV}}
\end{figure}

\begin{figure}[htbp]
\includegraphics[height=7cm,keepaspectratio=true]{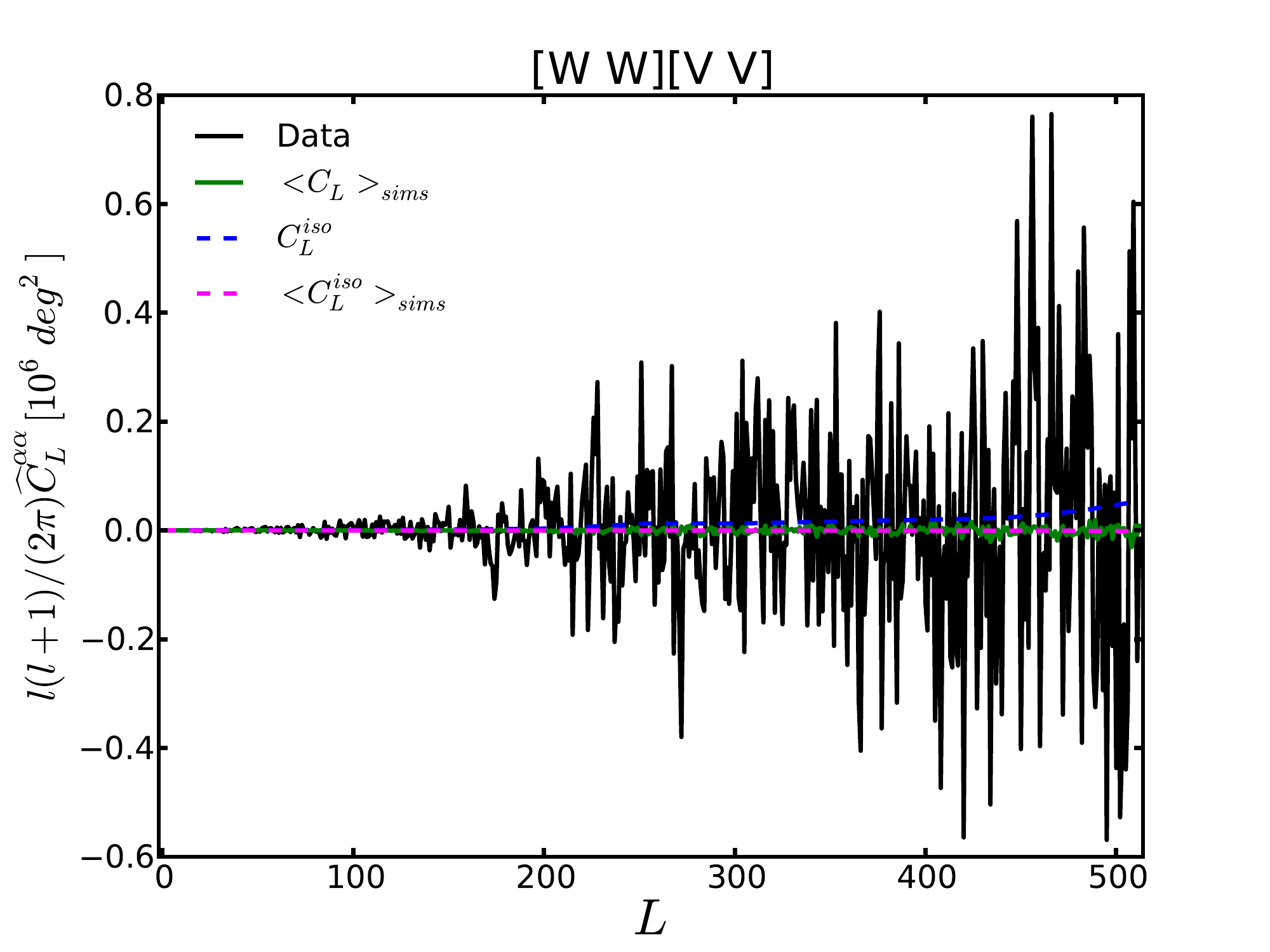}
\includegraphics[height=7cm,keepaspectratio=true]{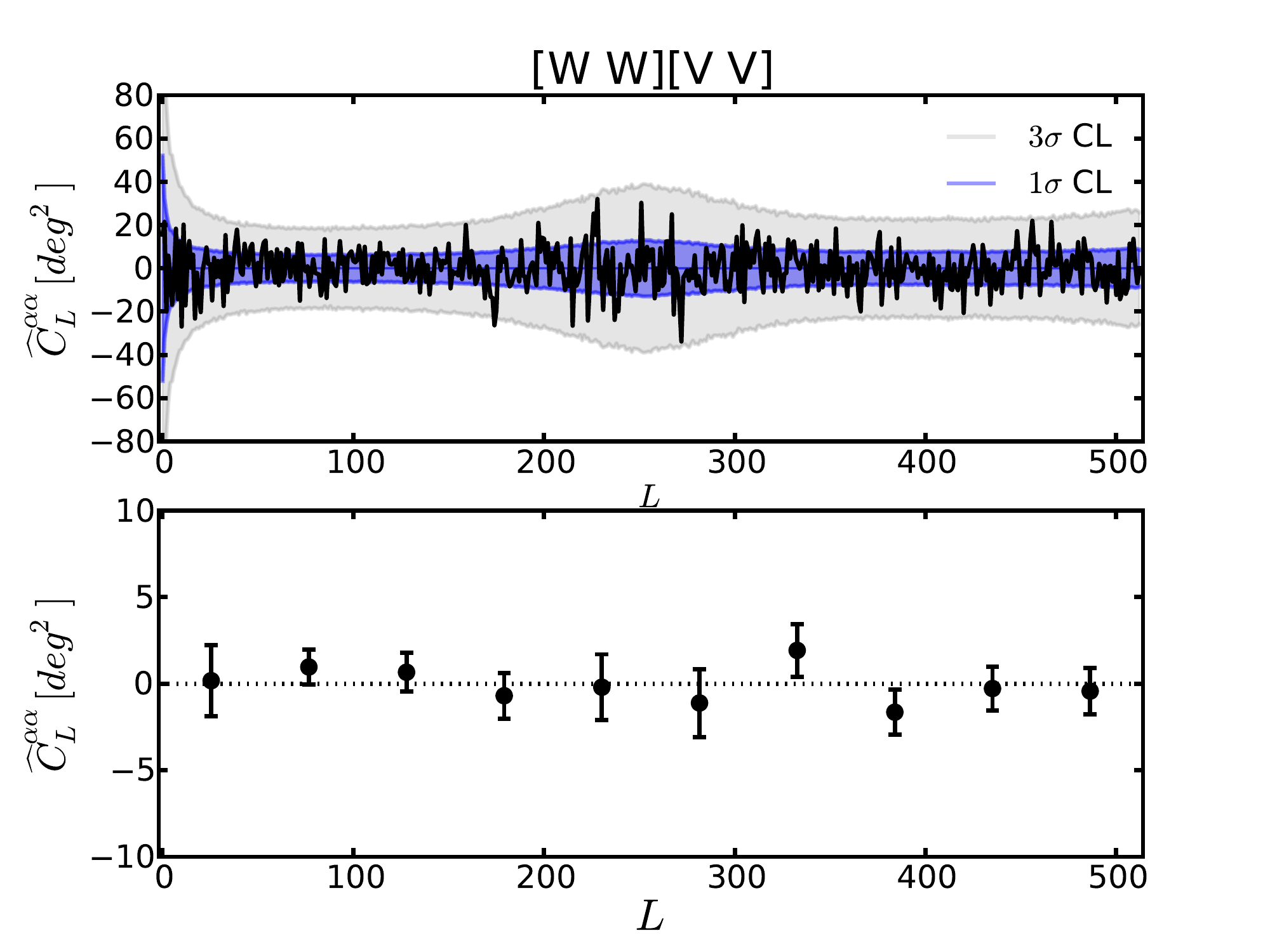}
\caption{Same as Figure \ref{fig:claa_VVVV}, for [$f_1f_2$][$f_3f_4$]=[WW][VV].\label{fig:claa_WWVV}}
\end{figure}

\begin{figure}[htbp]
\includegraphics[height=7cm,keepaspectratio=true]{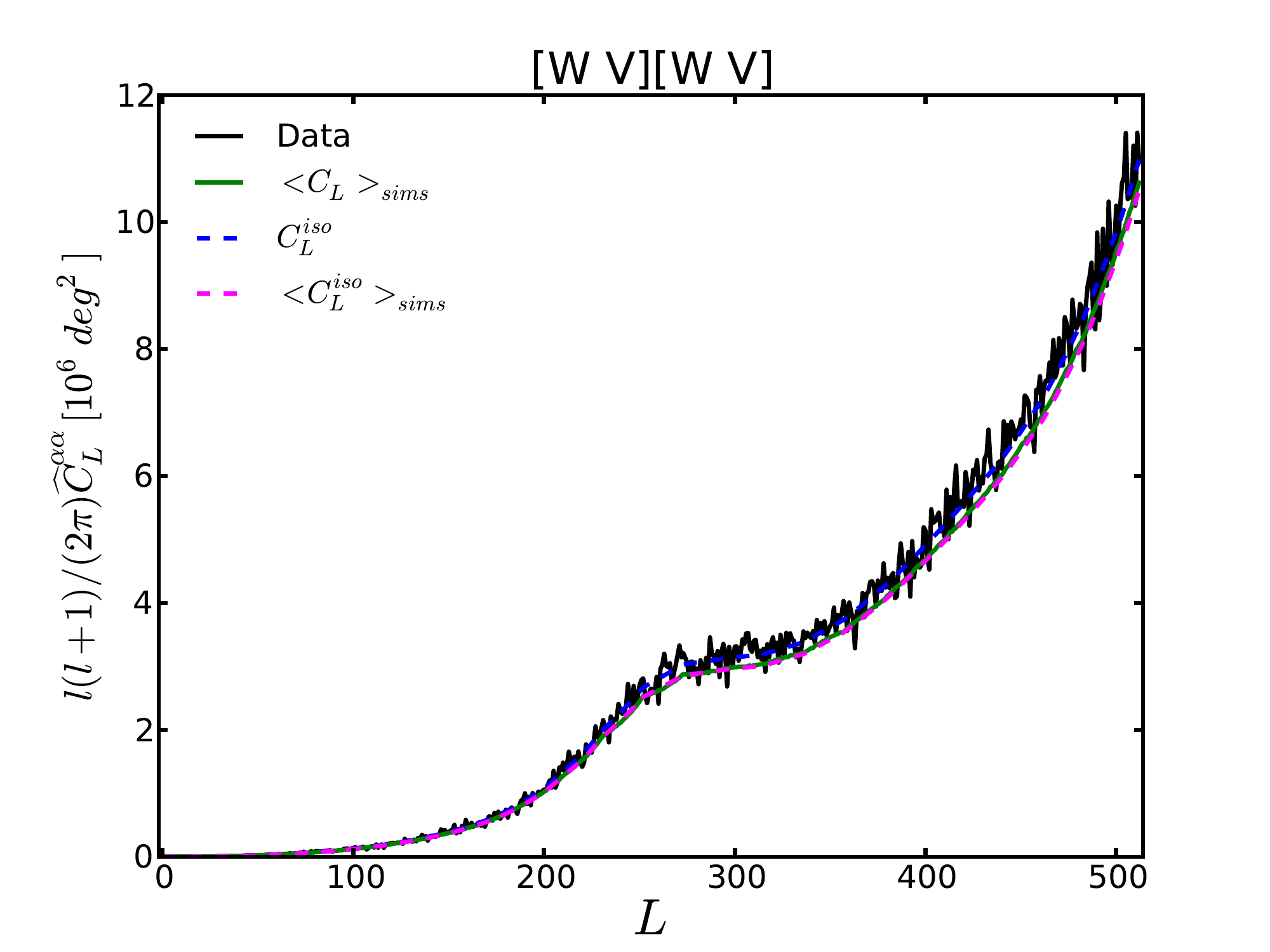}
\includegraphics[height=7cm,keepaspectratio=true]{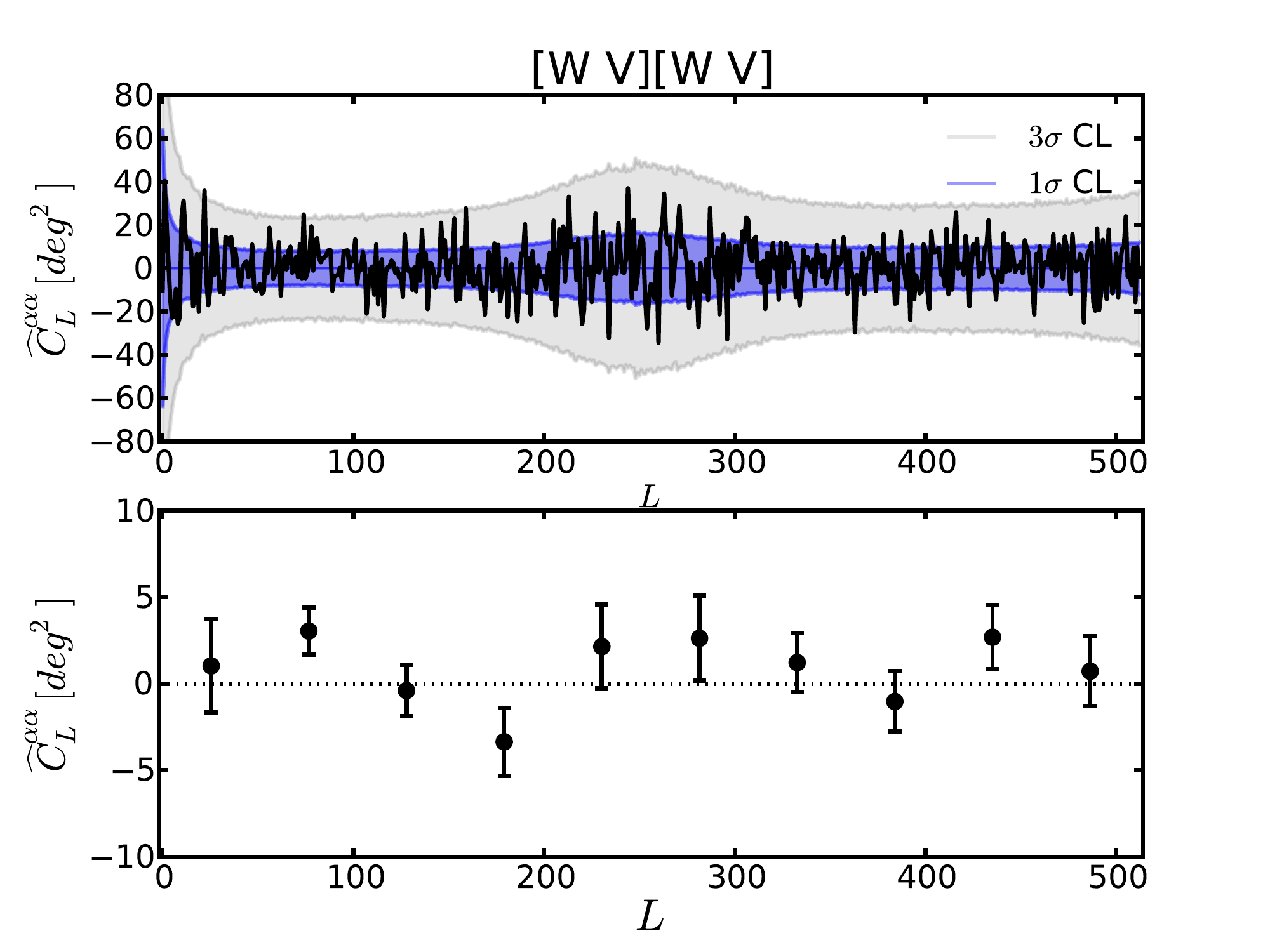}
\caption{Same as Figure \ref{fig:claa_VVVV}, for  [$f_1f_2$][$f_3f_4$]=[WV][WV].\label{fig:claa_WVWV}}
\end{figure}

Figs.~\ref{fig:claa_VVVV}, \ref{fig:claa_WWVV}, and  \ref{fig:claa_WVWV} show the measurement of the rotation-angle autocorrelation, before and after debiasing, and different components of the noise bias described in \S\ref{sec:caa_estimator}. The blue and gray areas in the middle panels represent $1\sigma$ and $3\sigma$ confidence-level intervals, respectively, derived from the null-hypothesis (no rotation) Monte Carlo analysis described in \S\ref{sec:simulations}. We see no significant deviations from zero in any of the five band cross-correlations---our results are consistent with $\alpha_{LM}=0$ to within $3\sigma$, at each multipole in the range from $L=0$ to $L=512$. We bin the power and list the measurements for all multipoles in Table \ref{tb:claa_all}. As an additional consistency check, the upper limit we obtain on the uniform rotation angle, given as $\alpha \equiv \alpha_{00}/\sqrt{4\pi}$, is in good agreement with previous WMAP results \cite{Komatsu:2010fb}; see Table \ref{tb:uniform}.

As we pointed out in \S\ref{sec:tests}, in the general case of arbitrarily large rotation, our method provides an exact estimate of the autocorrelation of the following quantity: $\frac{1}{2}\rm{sin}[2\alpha(\hatn)]$, rather than the rotation angle itself; when the small-angle approximation is satisfied, this quantity and its power spectrum assymptote to $\alpha(\hatn)$ and $C_L^{\alpha\alpha}$, respectively. In order to establish the regime corresponding to a particular model, we note that the RMS fluctuation typical of realization of a power spectrum $C_L^{\alpha \alpha}$ is given by
\beq
\langle \alpha(\hat{n})^2 \rangle^{1/2} = \sqrt{ \sum_{L} \frac{2L+1}{4\pi} C_L^{\alpha \alpha} }.
\label{eq:rms_claa}
\eeq
In the event of a breakdown in the small-angle approximation, the values in Table \ref{tb:claa_all} should be interpreted as constraints on the autocorrelation of $\frac{1}{2}\rm{sin}[2\alpha(\hatn)]$, rather than $\alpha$ itself. Evaluating Eq.~\eqref{eq:rms_claa} for the uncertainty levels quoted in Table \ref{tb:claa_all} would erroneously lead to a conclusion that a large RMS rotation is allowed by the WMAP data; we show that the upper limit on the RMS rotation is roughly $9.5^\circ$ (see Appendix \ref{ax:small_alpha}), and we again note that previous studies of quasar data imply an even stronger constraint of $\sim 4^\circ$ \cite{Kamionkowski:2010ss}.
\begin{table}[htbp]
\begin{center}
\begin{tabular}{ | c | c | c | c | c | c | p{10cm} |}
\hline
L bin & [VV][VV] & [QV][QV] & [QQ][VV] & [WV][WV] & [WW][VV] \\ \hline
26 & 2.65$\pm$1.87 & 1.61$\pm$2.44 & 1.05$\pm$1.62 & 0.72$\pm$2.03 & -0.43$\pm$1.34 \\ \hline
77 & 1.86$\pm$2.58 & 0.70$\pm$2.84 & 1.57$\pm$2.36 & 1.03$\pm$2.70 & 0.17$\pm$2.04 \\ \hline
128 & 1.07$\pm$1.33 & 1.00$\pm$1.36 & 0.27$\pm$1.17 & 3.04$\pm$1.35 & 0.96$\pm$1.02 \\ \hline
179 & 1.40$\pm$1.49 & -1.29$\pm$1.65 & -0.31$\pm$1.15 & -0.40$\pm$1.48 & 0.66$\pm$1.13 \\ \hline
230 & -1.90$\pm$1.76 & -4.47$\pm$1.96 & 1.87$\pm$1.36 & -3.36$\pm$1.97 & -0.69$\pm$1.33 \\ \hline
282 & 4.31$\pm$2.23 & 3.17$\pm$2.42 & 2.04$\pm$2.21 & 2.14$\pm$2.42 & -0.20$\pm$1.90 \\ \hline
333 & 1.98$\pm$2.39 & -0.25$\pm$2.60 & 4.59$\pm$1.80 & 2.62$\pm$2.45 & -1.11$\pm$1.96 \\ \hline
384 & 0.81$\pm$1.78 & -1.71$\pm$1.93 & 1.97$\pm$1.51 & 1.22$\pm$1.71 & 1.93$\pm$1.52 \\ \hline
435 & -0.40$\pm$1.64 & -0.19$\pm$1.80 & 1.53$\pm$1.26 & -1.03$\pm$1.74 & -1.65$\pm$1.30 \\ \hline
486 & 3.22$\pm$1.75 & 0.78$\pm$1.93 & 1.02$\pm$1.39 & 2.69$\pm$1.84 & -0.28$\pm$1.27 \\ \hline
\end{tabular}
\caption{Results for the measurement of $\widehat C_L^{\alpha\alpha}$ [degrees$^2$] are listed, as recovered from five different band cross-correlations. The $1\sigma$ confidence intervals are calculated with a suite of Gaussian sky simulations, described in \S\ref{sec:simulations}. The results are binned, with the central $L$ value of each bin listed in the table; the width of each bin is $\Delta L \sim 51$.}
\label{tb:claa_all}
\end{center}
\end{table}
\begin{table}[htbp]
\begin{center}
    \begin{tabular}{ | c | c | c | p{10cm} |}
    \hline
    [$f_1f_2$]& $\alpha \pm 1\sigma$ [${}^\circ$] \\ \hline
    [VV] & -0.9 $\pm$ 2.3\\ \hline
    [QV] & -0.5 $\pm$ 2.4\\ \hline
    [QQ] & 0.9 $\pm$ 2.8 \\ \hline
    [WV] & -2.2 $\pm$ 2.4\\ \hline
    [WW] & -1.8 $\pm$ 2.7\\ \hline
    \end{tabular}
\caption{Uniform-rotation angle $\alpha$ with a $1\sigma$ confidence interval, from five cross-band correlations of WMAP-7; the correction factor of $1/f_\text{sky}$ is applied to each measurement here. The uncertainties are consistent with the $\pm 1.4^{\circ}$ uncertainty on the uniform-rotation angle reported by the WMAP team \cite{Komatsu:2010fb} for a joint analysis of the Q, V and W-band data, after accounting for the fact that we analyze the bands individually (resulting in slightly larger error bars).}\label{tb:uniform}
\end{center}
\end{table}
\subsection{Constraints on a scale-invariant power spectrum}
\label{sec:SI_constraints}
The null result shown in \S\ref{sec:claa_constraints} can be translated into an upper limit on the amplitude of any model of rotation. As a generic example, we focus on a scale-invariant rotation-angle power spectrum of Eq.~\eqref{eq:caa_SI}.
The best-fit amplitude is evaluated from all multipoles in the range $0\leq L\leq 512$, using a minimum-variance estimator \cite{Gluscevic:2010vv}
\beq
\widehat A = \sigma_A^2 \sum_L \frac{C_L^{\alpha\alpha\text{,fiducial}}\widehat C_L^{\alpha\alpha}}{ \text{var}(\widehat C_L^{\alpha\alpha})},
\label{eq:Afit}
\eeq
where 
\beq
\sigma_A = \left(\sum_L \frac{(C_L^{\alpha\alpha\text{,fiducial}})^2}{\text{var} (\widehat C_L^{\alpha\alpha})}\right)^{-1/2},
\label{eq:sigmaA}
\eeq
is the analytic expression for the variance of $\widehat A$, and $\text{var}(\widehat C_L^{\alpha\alpha})$ is the variance of the null-hypothesis rotation-angle power spectrum, estimated from a suite of null-hypothesis simulations. We note that the measured $\widehat C_L^{\alpha\alpha}$ have been corrected by $f_\text{sky}(L)$ (see \S\ref{sec:tests} and Appendix \ref{ax:Lfsky}; the correction is calculated specifically for this model) only in this Subsection; for the presentation of the model-independent results, we use the average value, $f_\text{sky}\sim 0.68$. Most of the constraint here comes from low $L$'s; $50\%$ of the sum in Eq.~\eqref{eq:sigmaA} comes from $L=1$, and $90\%$ from $L<10$. 

\begin{figure*}
\includegraphics[height=4cm,keepaspectratio=true]{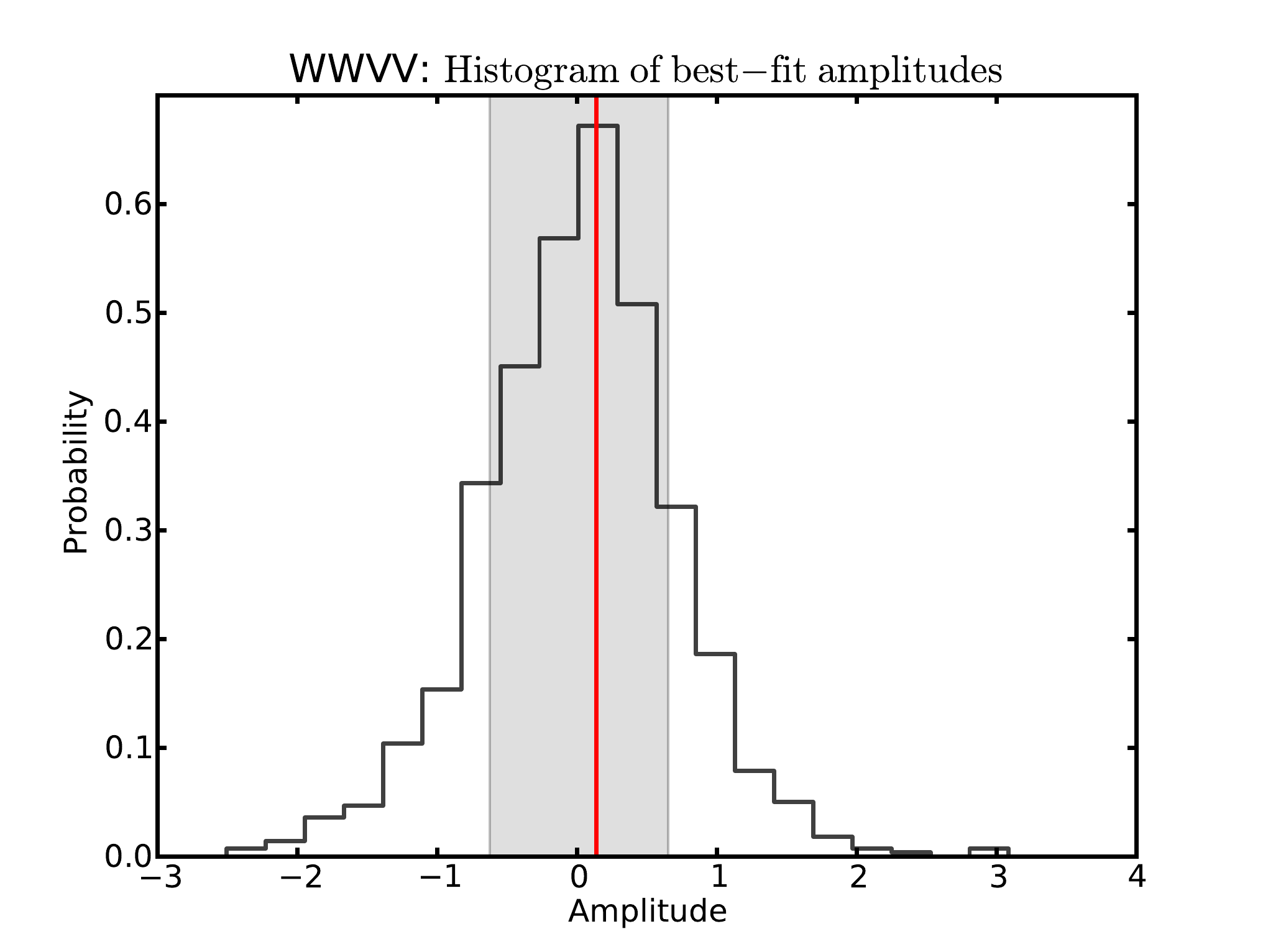}
\includegraphics[height=4cm,keepaspectratio=true]{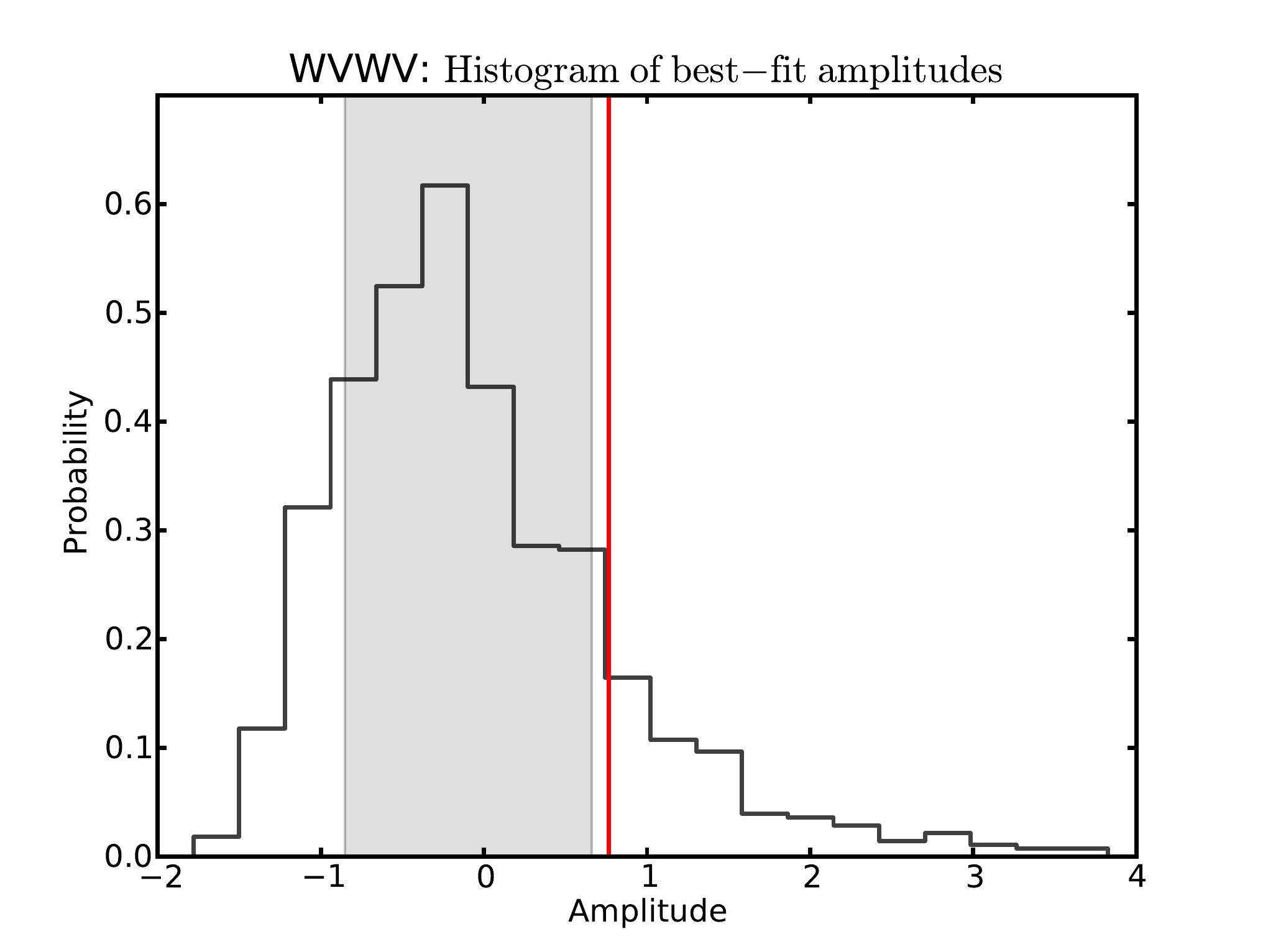}
\includegraphics[height=4cm,keepaspectratio=true]{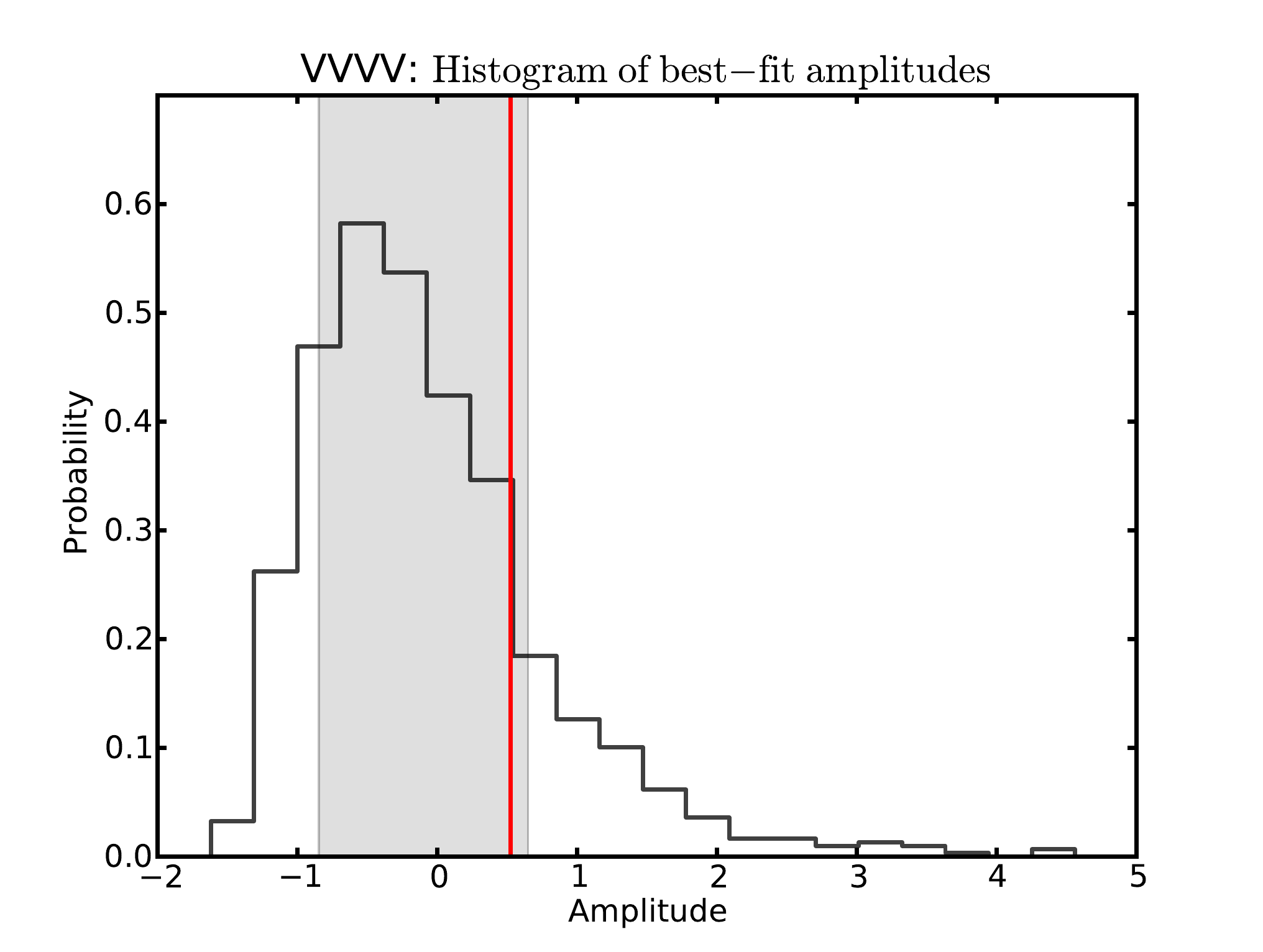}
\caption{Probability distributions of the best-fit amplitude $\widehat A$ of the scale-invariant rotation power $C_L^{\alpha\alpha}$ recovered from  a suite of null-hypothesis realizations are shown for some of the cross-band correlations. The gray-shaded areas denote a $68\%$ confidence interval around a mean value; the red vertical line represents the measurement of the best-fit $\widehat A$ for a given band-cross correlation. We find consistency with zero within $3\sigma$ for all five measurements.
\label{fig:Ahistograms}}
\end{figure*}

Even though the analytic expression above provides a good estimate of the statistical variance, because the constraint comes primarily from low-L modes the probability distribution of $\hat{A}$ is significantly non-Gaussian. To capture this non-Gaussianity in our analysis, we again generate a suite of null-hypothesis Monte Carlo simulations and recover the $68\%$ and $99\%$ confidence-level intervals from these simulations. The corresponding probability distributions for $\widehat A$ are shown in Figure \ref{fig:Ahistograms}. 

The best-fit values for the quadrupole amplitude $\widehat C_2^{\alpha\alpha}$ and associated confidence intervals are listed in Table \ref{tb:amplitude}; consistency with zero is apparent within $3\sigma$  for all band-cross correlations we analyzed. The tightest constraint on the quadrupole amplitude of a
scale-invariant rotation-angle power spectrum comes from
[WW][VV]; it is $\sqrt{C_2^{\alpha\alpha}/(4\pi)}\lesssim 1^{\circ}$ with $68\%$ confidence\footnote{Note that the conversion between the amplitude $A$ and the quadrupole is $C_2 = A \times 131 \text{deg}^2 / 6$.}.
\begin{table}[htbp]
\begin{center}
    \begin{tabular}{ | c | c | c | p{10cm} |}
    \hline
    [$f_1f_2$][$f_3f_4$] & $\widehat C_2^{\alpha\alpha}\pm 1\sigma (\pm 3\sigma)$ [deg$^2$] \\ \hline
    [VV][VV] &  $11.4^{+15.8}_{-16.9}  (^{+79.0}_{-27.7})$  \\ \hline
    [QV][QV] &  $29.6^{+18.8}_{-18.3}  (^{+70.3}_{-33.4})$  \\ \hline
    [QQ][VV] &  $19.8^{+14.3}_{-13.9} (^{+51.6}_{-46.6})$  \\ \hline
    [WV][WV] &  $16.8^{+15.9}_{-16.9} (^{+79.0}_{-27.7})$  \\\hline
    [WW][VV] &  $3.0^{+14.0}_{-13.9}  (^{+43.3}_{-42.9})$  \\ \hline
    \end{tabular}
\caption{Measurement of the quadrupole amplitude of a scale-invariant rotation-angle power spectrum for different cross-band correlations, with $68\%$ and $99\%$ confidence-level intervals, recovered from a suite of null-hypothesis simulations. 
Consistency with zero within $3\sigma$ is apparent for all band cross-correlations, and the tightest constraint comes from [WW][VV].}\label{tb:amplitude}
\end{center}
\end{table}
\subsection{Potential systematics}
\label{sec:systematics}
In addition to the statistical error reported here, there is also a systematic error for the measurement of the uniform rotation angle, owing to uncertainty in the detector polarization angles \cite{Komatsu:2010fb}. This systematic uncertainty should only apply to the monopole of $\alpha$. The direction-dependent part is only sensitive to the extent that it affects the statistical noise bias, and this is mitigated by our data-dependent debiasing procedure. There are, however, other sources of systematic error that can potentially bias our estimates and add uncertainties to the rotation-angle measurements. In this Section we investigate the impact of asymmetry of the instrumental beams, unresolved polarized point sources, and foreground residuals from unremoved/unmasked Galactic emission. 
\subsubsection{Beam Asymmetries}
\label{sec:beams}
The fast spin and precession rates of the WMAP scan strategy, coupled with the yearly motion of the satellite around the ecliptic, enforce that any bias to $\widehat\alpha_{LM}$ originating from scan-strategy related systematics, like beam asymmetry, must be confined to $M=0$ modes in ecliptic coordinates \cite{Hanson:2010gu}. Furthermore, the smoothness of the scan strategy on large scales (dictated by the $85$-degree opening angle of the detectors, and the large $22.5$-degree amplitude of the hourly satellite precession), ensure that any such bias falls off quickly as a function of $L$. The estimates of $C_2$, as we have discussed in the previous section, are most sensitive to the low-L modes of $\widehat C^{\alpha\alpha}_L$, so to test for the presence of beam-asymmetry contamination, we rotate our coordinate system to ecliptic coordinates, and re-derive a constraint on $C_2$ from $L<10$, $M=0$ modes. We see no departure from the null hypothesis in this case where it should be maximal, and so conclude that beam asymmetries are not a significant source of bias for our measurements. 
\subsubsection{Unresolved point sources}
\label{sec:point_sources}
\begin{figure}[htbp]
\includegraphics[height=7cm,keepaspectratio=true]{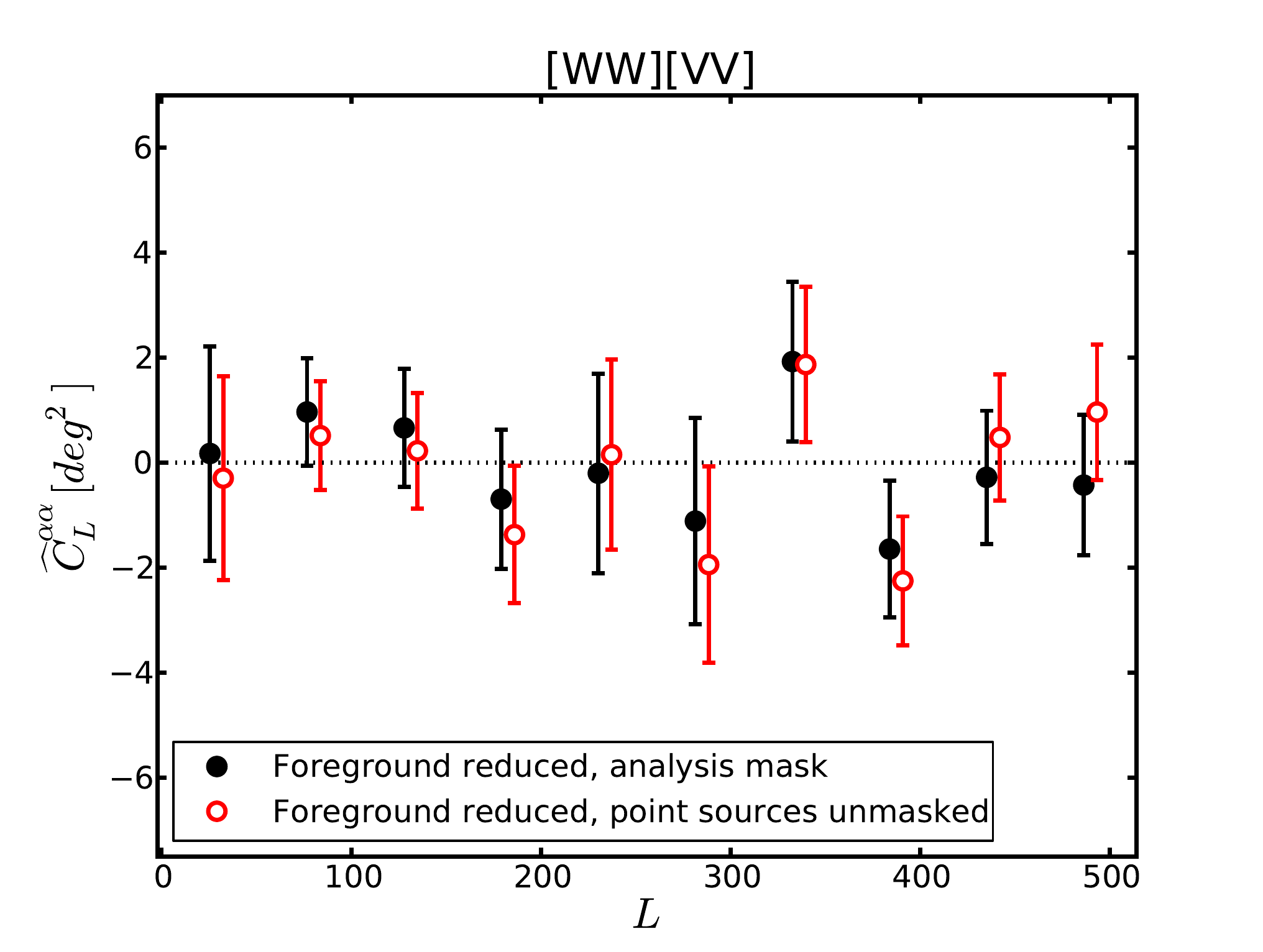}
\caption{Measurement of $\widehat C_L^{\alpha\alpha}$ from [WW][VV]. Results shown in black are obtained by using the analysis mask that covers all the point sources brighter than $\sim 1$Jy, while the results in red (empty circles) are obtained after unmasking all the point sources. There is no apparent bias and the difference in every bin is less than the statistical uncertainty, despite the extreme variation in the source contamination. Scaling arguments in \S\ref{sec:point_sources} imply that the unresolved point sources have a negligible contribution to the estimated measurement uncertainty for the most constraining band-cross correlation.
\label{fig:points_comparison}}
\end{figure}

To test the impact of unresolved point sources on our results, we repeat the analysis after unmasking the portion of the maps which are associated with detected point sources (but not Galactic contamination). Compared to our fiducial analysis, the measurement points for $\widehat C_L^{\alpha\alpha}$ shift by $\lesssim 1\sigma$, where $\sigma$ represents the statistical error from our foreground-free Monte Carlo analysis; see Figure \ref{fig:points_comparison}.  This shift provides a conservative upper limit on the systematic uncertainty that point-source residuals can produce, assuming that the bright detected population has comparable polarization properties to those of the fainter sources. The unresolved point-source power at WMAP frequencies is dominated by unclustered radio sources, with fluxes close to the detection threshold, and so this is a reasonable assumption. We note that there is no overall bias, as the direction of scatter does not appear to be correlated for different multipoles. Of course, the contribution of radio point sources to the map is a steep function of the flux cut, and by unmasking all detected point sources our estimate of potential bias and uncertainty is overly conservative. Given a model for radio-source number counts $dN/dS$, we can scale these results to the levels of contamination expected at the actual WMAP source detection threshold of (conservatively) $\sim 1$~Jy. Any bias $\Delta \widehat{C}_L^{\alpha \alpha}$ (which we do not see evidence for, even in the unmasked map) will scale with the point-source trispectrum as
\beq
\Delta \widehat{C}_L^{\alpha \alpha} \propto \int_{S=0}^{S_\text{cut}} S^4 \frac{dN}{dS} dS,
\eeq
while the uncertainty on $\widehat{C}_L^{\alpha\alpha}$ will scale with the point-source power as
\beq
\sigma \left( \widehat{C}_L^{\alpha \alpha} \right) \propto \left( \int_{S=0}^{S_\text{cut}} S^2 \frac{dN}{dS} dS \right)^2.
\eeq
Evaluating these terms for the $dN/dS$ model of Ref.~\cite{DeZotti:2009an}, we find that $\Delta$ and $\sigma$ should be suppressed by factors of $0.005$ and $0.06$ respectively when moving from a flux cut of $10$~Jy (no masking) to $1$~Jy. We find even smaller (though comparable in magnitude) results using the simpler ${dN}/{dS} \propto S^{-2.15}$ model of Ref.~\cite{Bolton:2004tw}. This implies that any bias from unresolved sources should be completely negligible, and any increase in uncertainty due to their contribution to the observed power should be $\lesssim 0.1\sigma$, where $\sigma$ represents the statistical error from our point-source--free Monte Carlo analysis. 
In conclusion, we expect the unresolved point sources to produce a negligible systematic uncertainty in the measurement of $\widehat C_L^{\alpha\alpha}$.
\subsubsection{Foreground residuals}
\begin{figure}[htbp]
\includegraphics[height=7cm,keepaspectratio=true]{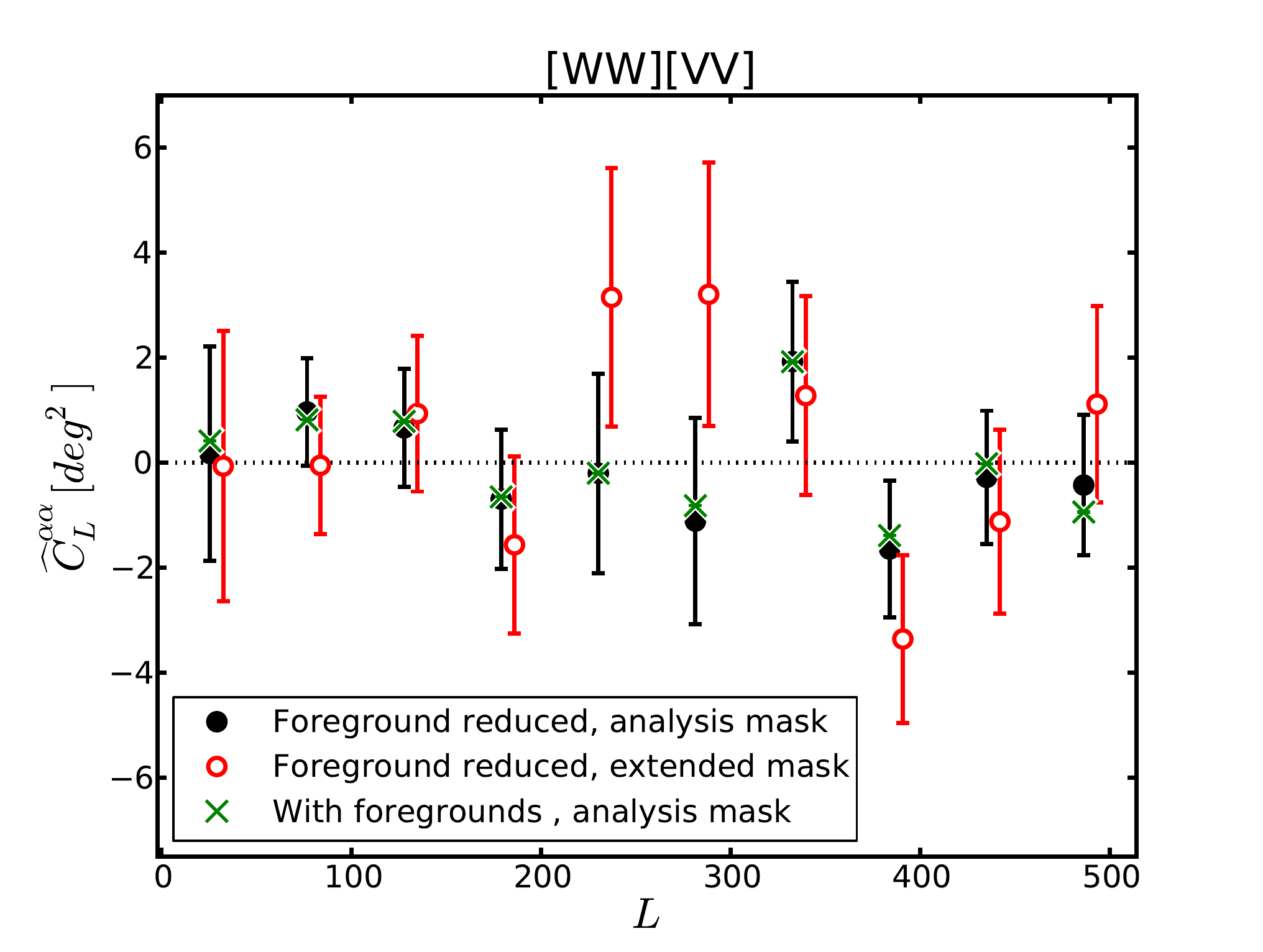}
\caption{Measurement of $\widehat C_L^{\alpha\alpha}$ from [WW][VV] band cross-correlation, with the corresponding statistical uncertainty obtained from a suite of null-hypothesis foreground-free simulations. Black filled circles are measurements obtained from the foreground-reduced maps after applying the fiducial analysis mask, and they represent our fiducial results of Figure \ref{fig:claa_WWVV}. The colored data points (and the associated error bars) correspond to the two test analyses: the green $x$'s are obtained from the maps prior to foreground subtraction, but using the fiducial mask, while the red empty circles are measurements obtained from foreground-reduced maps after applying an extended mask (with an additional $\sim 35\%$ of the sky covered around the Galactic plane). The results from the two tests show no apparent bias. For the case of foreground-non-reduced analysis, the difference from the fiducial measurements is negligible compared to the statistical uncertainty; for the extended-mask case, the scatter between the two results is consistent with the difference in sky coverage. This implies that foregrounds and foreground residuals should not have a drastic impact on the estimated measurement uncertainty.
\label{fig:fg_comparison}}
\end{figure}

An additional conceivable source of systematic uncertainty might result from Galactic foregrounds.  To explore the extent to which such uncertainty might affect our results, we perform two tests.  In the first, we repeat our analysis on non-foreground-reduced maps, to test the effect of the presence of unsubtracted foregrounds.  In the second, we repeat the foreground-reduced analysis using a mask which excludes a larger fraction of the low-Galactic-latitude sky. We construct this conservative mask by combining the fiducial $KQ85y7$ analysis mask with the extended mask of Ref.~\cite{lambda}, and additionally masking out pixels with Galactic latitudes in the range of $\pm 40^\circ$; the mask admits only about $33\%$ of the sky, approximately half the sky admitted in our fiducial analysis (where $\sim 68\%$ of the sky is admitted; see Appendix \ref{ax:masks}). The function of this test is to explore the effect of residuals left by the foreground subtraction procedure, which should be stronger close to the Galactic plane.  The measurement of $\widehat C_L^{\alpha\alpha}$ is scaled appropriately to account for the fractional sky coverage and the results from the two modified analyses are compared with the results of the fiducial analysis in Figure \ref{fig:fg_comparison}. In the first case, the change in the measurements is negligible compared to the statistical uncertainty. In the second case, the scatter between the two results is consistent with the difference in sky coverage (producing up to $40\%$ larger scatter for the extended-mask data points). The measurements show no apparent bias in either case. These results imply that foregrounds and foreground residuals are not likely to make a large systematic contribution to our estimated statistical uncertainty, at least for the case of the most constraining band-cross correlation [WW][VV]. 
\section{Summary and conclusions}
\label{sec:discussion}
In this work, we implement the minimum-variance quadratic-estimator
formalism of Ref.~\cite{Gluscevic:2009mm} to search for direction-dependent cosmological
birefringence with WMAP 7-year data. We derive the first CMB
measurement of the rotation-angle power spectrum in the range
$L=0-512$, finding consistency with zero at each multipole,
within $3\sigma$. We estimate an upper limit on
each power-spectrum multipole by simulating a suite of Gaussian sky
realizations with no rotation, including symmetric beams and
noise realizations appropriate for each WMAP frequency band, and
also $Q$-$U$ correlations and sky cuts. We investigate the impact of foregrounds and polarized diffuse point sources on the reported constraints and come to the conclusion that they are not significant sources of systematic error for the rotation-angle estimates. Finally, we use the null result to get a $68\%$ confidence-level upper limit of $\sqrt{C_2^{\alpha\alpha}/(4\pi)}\lesssim 1^{\circ}$ on the quadrupole of a scale-independent rotation-angle power spectrum. Even though the CMB constraint turns out to be comparable to that derived from quasar measurements, the CMB analysis has a significant advantage: it provides a measurement of the rotation-angle power at each individual multipole $L$ and has better sensitivity to models with significant power at high multipoles. 

The same formalism we use in this work can be applied to forecast the sensitivity of upcoming and future-generation CMB satellites to detecting direction-dependent cosmological birefringence. With 7 years worth of integration time with WMAP, we are able to constrain the uniform component of the rotation to less than about a degree; it will be interesting to see the results of this analysis method applied to the upcoming data from Planck satellite \cite{Planck}, where the sensitivity to rotation angles on the order of a few arcminutes \cite{Gluscevic:2009mm} is expected. 
\begin{acknowledgments}
This work was supported
by DoE DE-FG03-92-ER40701, NASA NNX12AE86G, DoE (DOE.DE-SC0006624), and the David and Lucile Packard Foundation. Part of the
research described in this paper was carried out at the Jet
Propulsion Laboratory, California Institute of Technology, under
a contract with the National Aeronautics and Space
Administration. Some of the results in this paper have been
derived using the HEALPix package \cite{Gorski:2004by}.
\end{acknowledgments}
\appendix
\section{Constraints on rms rotation from WMAP}
\label{ax:small_alpha} 
\begin{figure}[htbp]
\includegraphics[height=7cm,keepaspectratio=true]{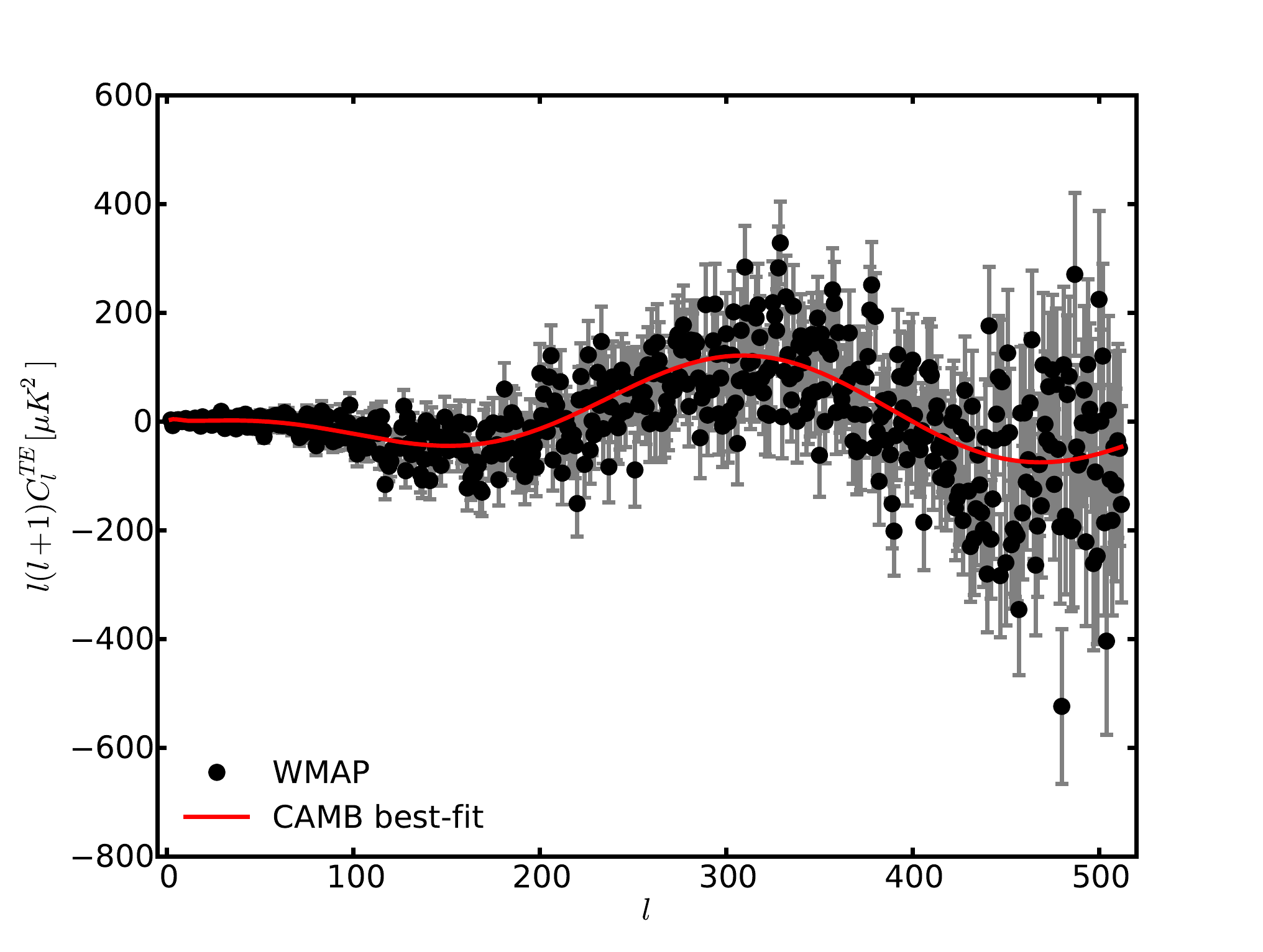}
\caption{$TE$ correlation measured from WMAP-7 data (black) is compared to the primordial power spectrum, generated using CAMB for the best-fit cosmology with no rotation (red, solid); both power spectra are obtained from Ref.~\cite{lambda}. The uncertainty on this measurement (gray) leaves room for a maximal rotation-angle rms of roughly $9.5^\circ$.\label{fig:clte_wmap7}}
\end{figure}

If the primordial B mode is small compared to the primordial E mode, and the rotation field is independent of the CMB, the measured $TE$ correlation reads (see also Figure \ref{fig:clte_wmap7})
\beq
C_l^{TE} = \left< \rm{cos}[2\alpha(\hatn)]\right> \widetilde C_l^{TE},
\label{eq:clte_cos2a}
\eeq
where the mean is taken over all realizations of the rotation field, and it does not depend on the direction $\hatn$. In the case the probability distribution for $\alpha$ is a Gaussian centered at zero and with a width $\left<\alpha(\hatn)^2\right>^{1/2}$, the expectation value in Eq.~\eqref{eq:clte_cos2a} is simply related to the pixel-variance of $\alpha$,
\beq
\left< \rm{cos}[2\alpha(\hatn)]\right> = e^{-2\left<\alpha(\hatn)^2\right>}.
\label{eq:expectation_cos2a}
\eeq
Therefore, an estimate of this expectation value and its uncertainty, obtained from the $TE$ measurement as compared to the primordial power sectrum $\widehat C_l^{TE}$ provides an upper limit of the rotation-angle pixel-variance. Adopting the expressions for a minimum-variance estimator and its variance (analogous to those of Eqs.~\eqref{eq:Afit} and \eqref{eq:sigmaA}), we obtain: $\left< \rm{cos}[2\alpha(\hatn)]\right> = 0.997 \pm 0.050$ (note that this constraint follows from the $21\sigma$ confidence-level detection of the $TE$ correlation reported by Ref.~\cite{Komatsu:2010fb}), implying an upper limit on the rotation rms $\left<\alpha(\hatn)^2\right>^{1/2}\lesssim 9.5^\circ$. 
\section{Visualizing the CB kernels}
\label{ax:kernels} 
\begin{figure*}[htbp]
\includegraphics[height=4cm,keepaspectratio=true]{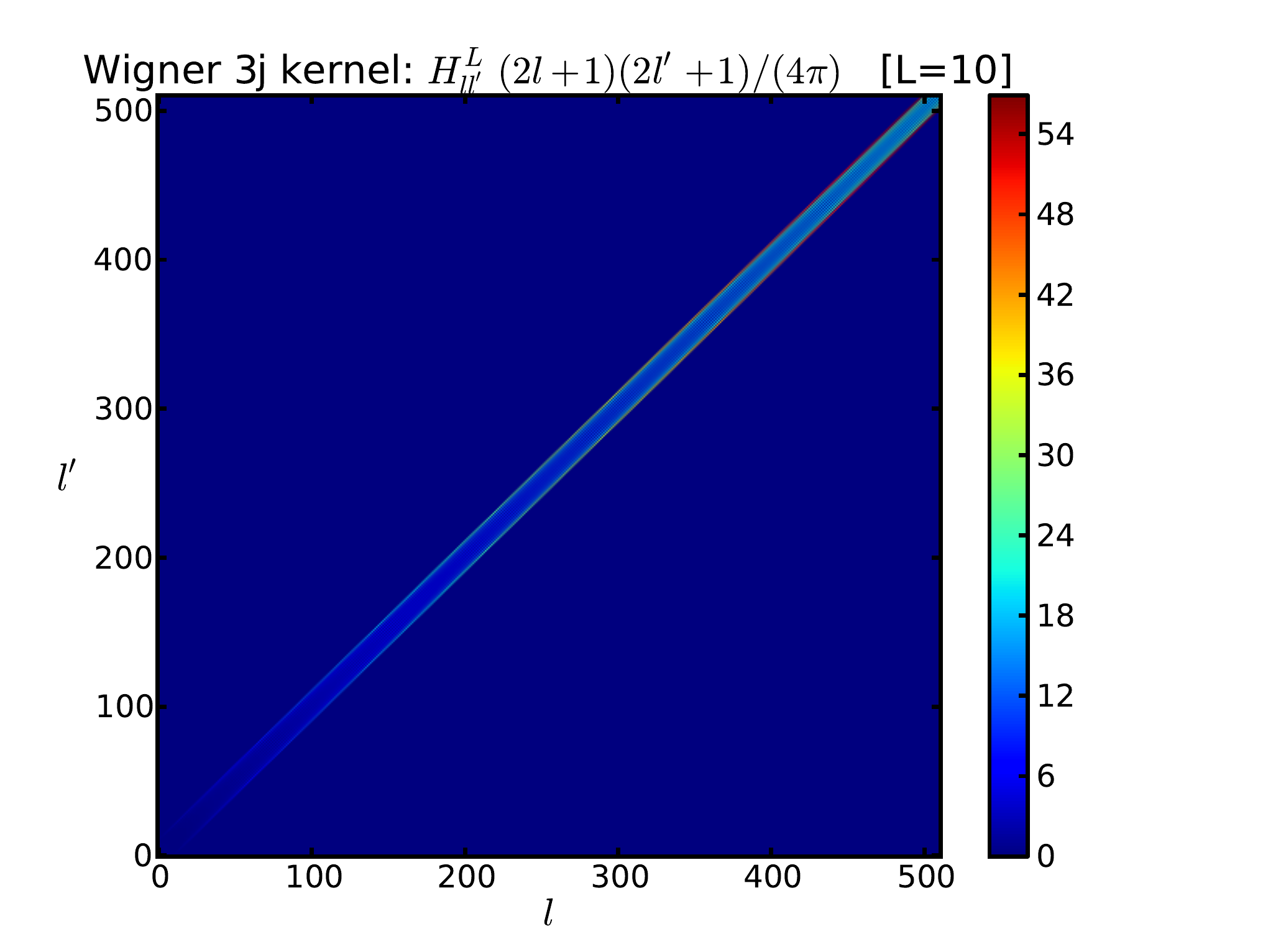}
\includegraphics[height=4cm,keepaspectratio=true]{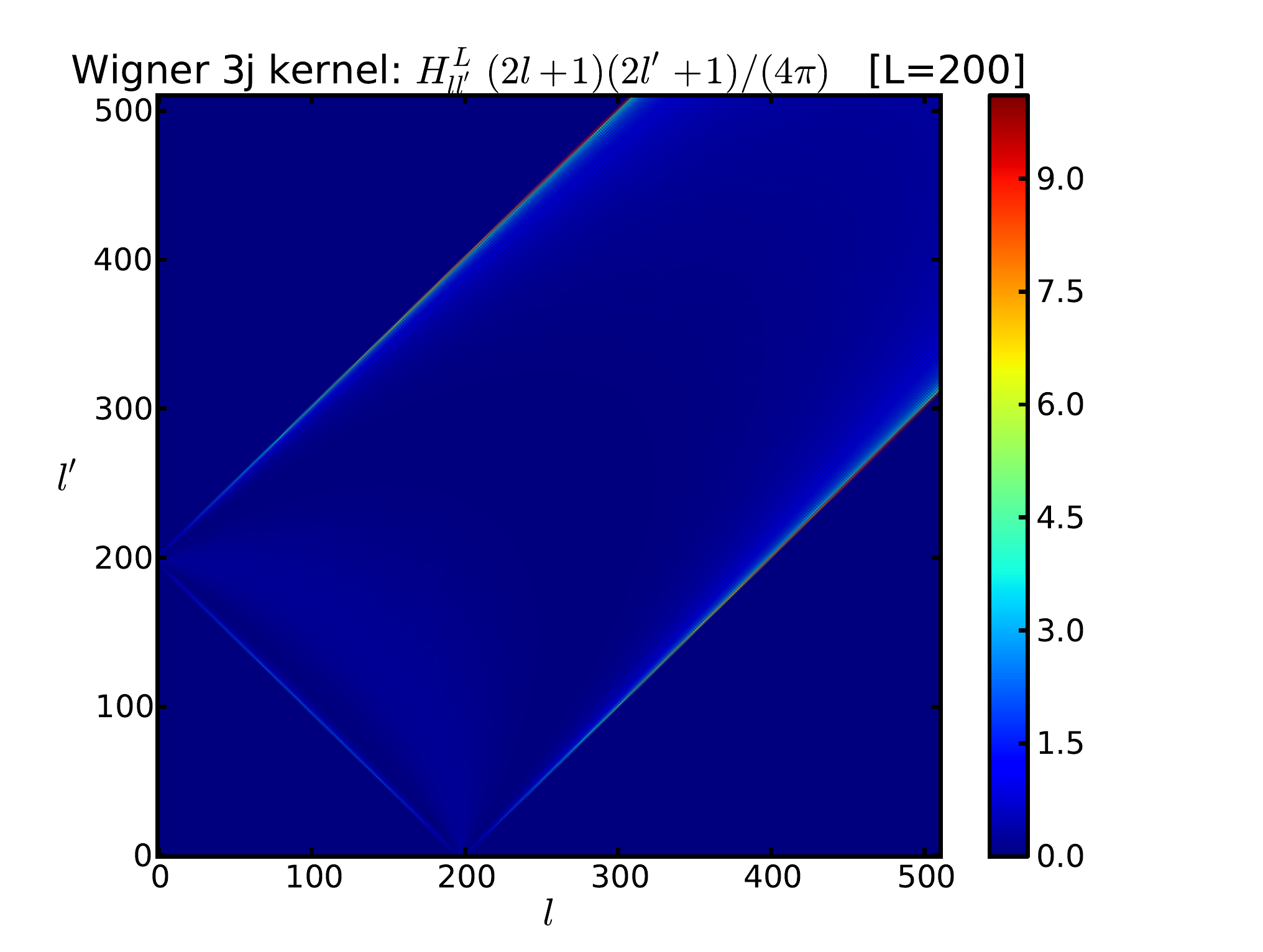}
\includegraphics[height=4cm,keepaspectratio=true]{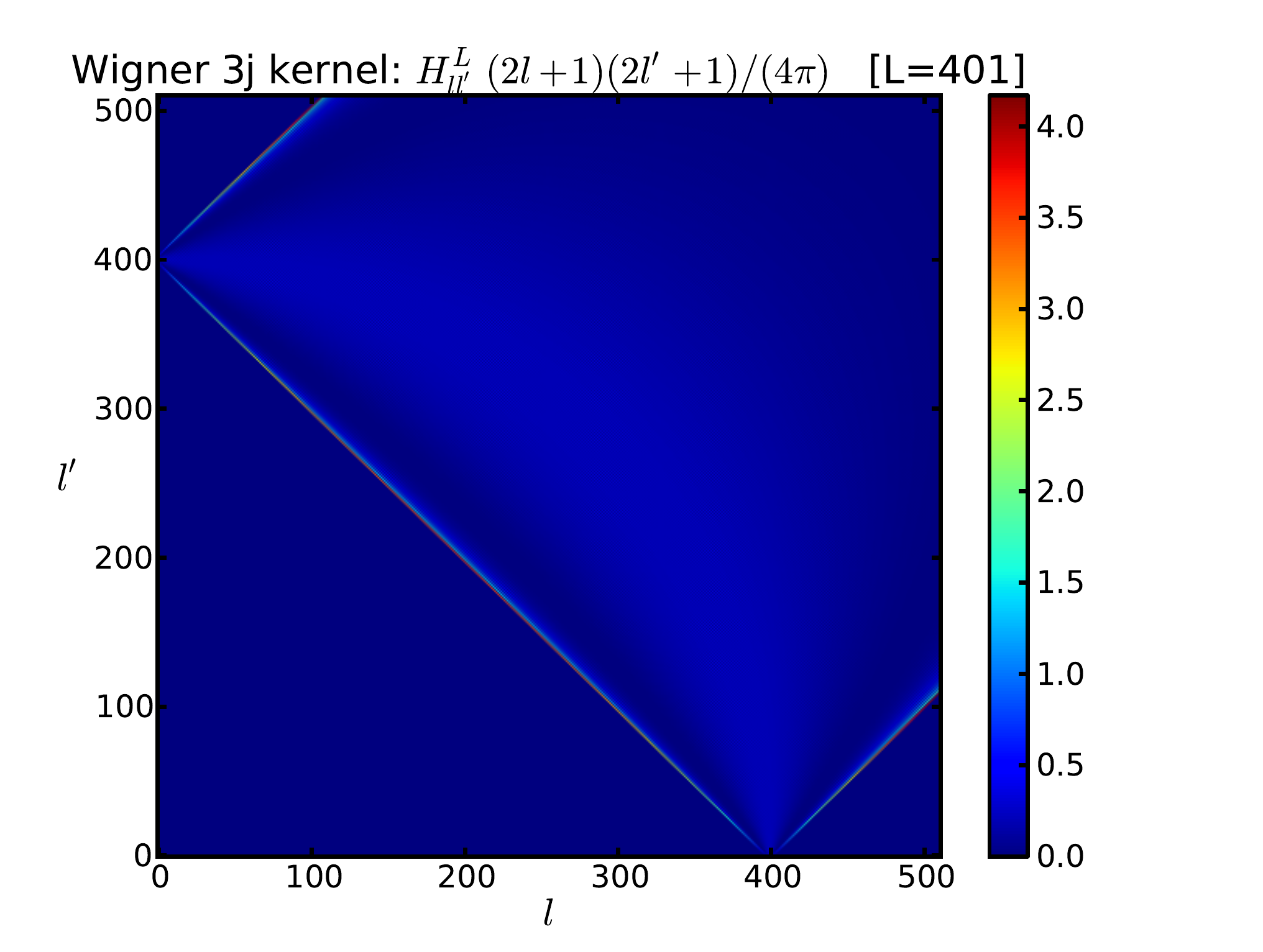}
\caption{The Wigner-3j geometric factors in the summands of Eqs.~\eqref{eq:estimator} and \eqref{eq:iso_bias} for three different values of $L$ are shown. The geometric factor is non-zero only in the region of the $ll'$ space where the triangle inequalities are satisfied. The dominant contribution comes from the triangles in which $L\sim l$, or $L\sim l'$, i.e. where either the temperature or the polarization mode has a length scale comparable to the length scale of the rotation-angle mode.\label{fig:w3j_kernel}}
\end{figure*}

\begin{figure}[htbp]
\includegraphics[height=7cm,keepaspectratio=true]{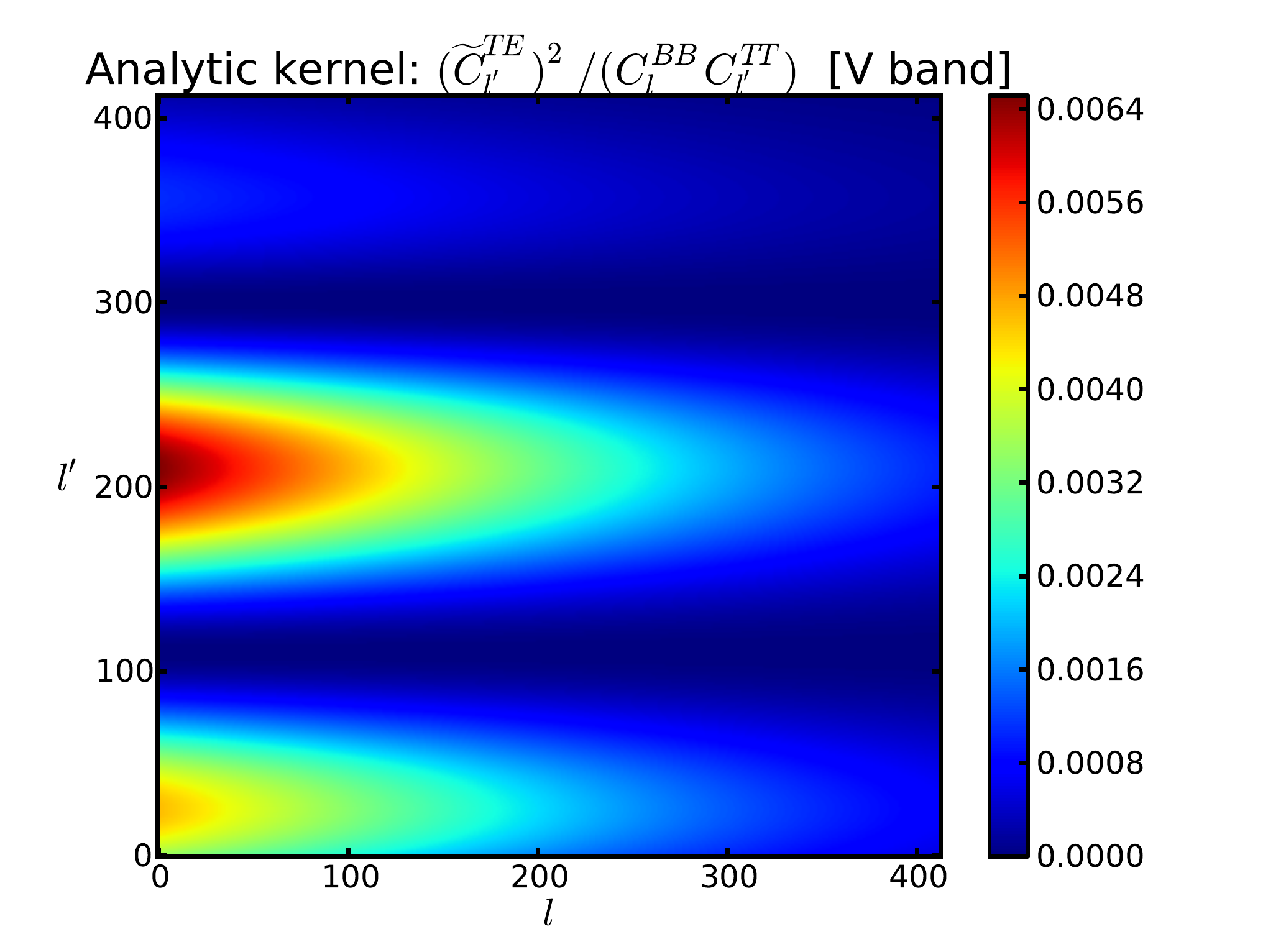}
\caption{The power-spectrum kernel of the summands in Eqs.~\eqref{eq:estimator} and \eqref{eq:iso_bias} is shown. The types of triangles that contribute the most to the isotropic bias of $C_L^{\widehat\alpha\widehat\alpha}$ are set by the geometric properties of spin-2 Wigner-3j symbols illustrated by the kernel shown in Figure \ref{fig:w3j_kernel}, which is modulated by this kernel to produce summands in the expression for the bias.\label{fig:cl_kernel}}
\end{figure}

To illustrate the shape in harmonic space of the statistical
anisotropy introduced by CB, we plot here the power-spectrum
kernel as well as the geometric Wigner-3j contributions to the
CB kernels in the $ll'$ space from Eq.~\eqref{eq:estimator}; see
Figs.~\ref{fig:w3j_kernel} and \ref{fig:cl_kernel}. The
structure of the power-spectrum kernel originates from the
polarization and temperature power spectra; the terms that
correspond to the acoustic peaks in the primordial $TT$ and $TE$
power spectra have the largest contribution to the sum over $ll'$. The
geometric weight dictates the shape of the $l, l', L$ triangles
which are generated by CB at a scale $L$.  The terms where
either $l$ or $l'$ is close in value to $L$ have the
largest contribution. The combination of the geometric weight and the
power-spectra weight dictates the size of the noise bias at any
particular $L$. The interplay of the two, for example, produces
a peak at $L\sim 270$, apparent in all the plots of the noise
bias presented in this work.  The local maximum in the variance of $\widehat
C_L^{\alpha\alpha}$ also appears at this scale.
\section{$f_\text{sky}(L)$}
\label{ax:Lfsky}
\begin{figure}[htbp]
\includegraphics[height=7cm,keepaspectratio=true]{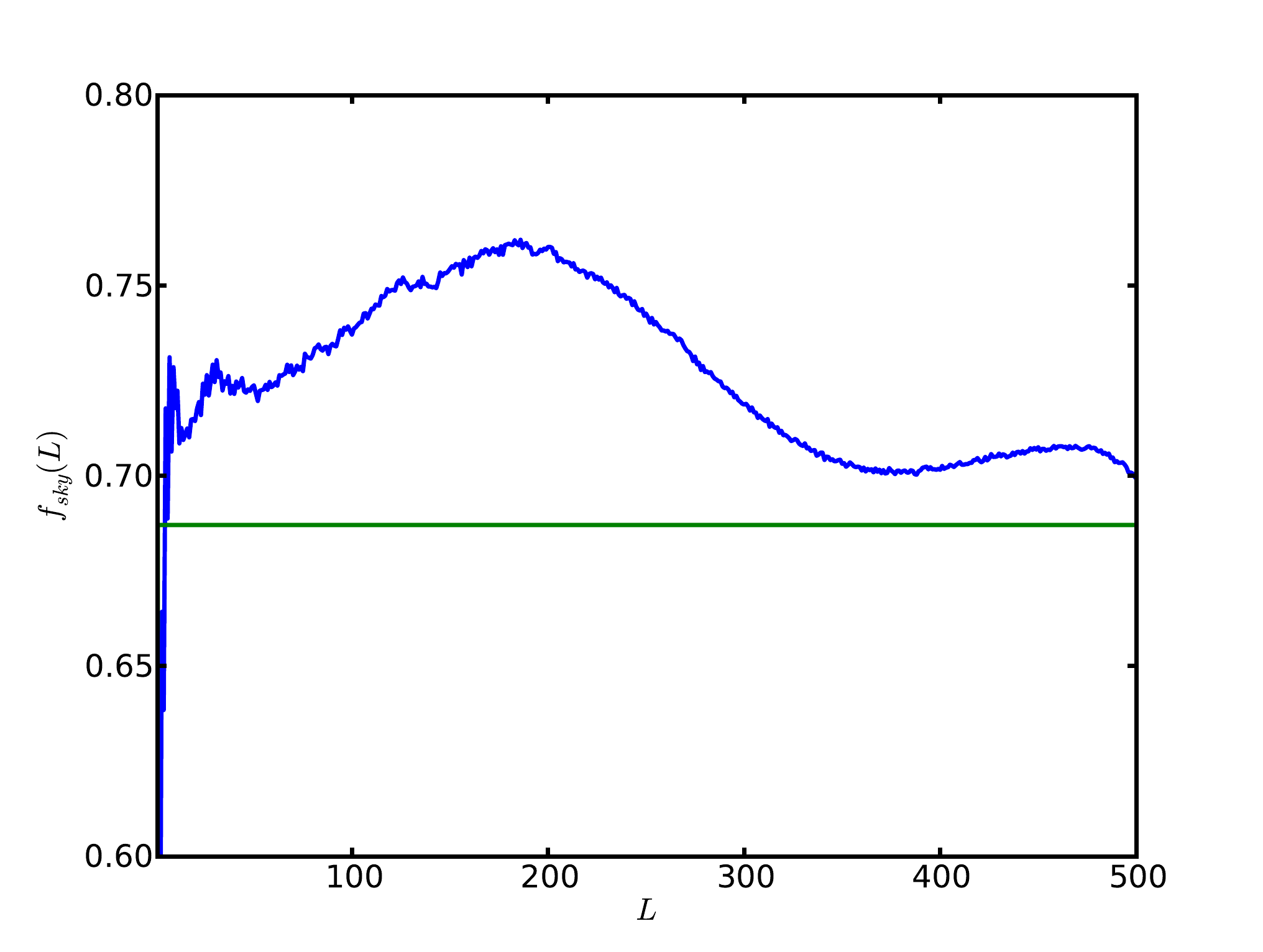}
\caption{The $L$ dependence of the $f_\text{sky}$ factor used for the reconstruction of the scale-invariant rotation-angle power spectrum in \S\ref{sec:tests} and \S\ref{sec:SI_constraints}. The horizontal (green) line at $f_\text{sky}=0.68$ represents the fraction of pixels admitted by the mask.
\label{fig:Lfsky}}
\end{figure}

In order to evaluate the exact $L$-dependence of $f_\text{sky}$ used to reconstruct the scale-invariant rotation-angle power from the CMB maps (see \S\ref{sec:tests}), we generate a large number of $\alpha(\hatn)$ realizations of the power-spectrum model of Eq.~\eqref{eq:caa_SI}, mask the sky with the fiducial analysis mask, and then recover the input power spectrum in the usual way, i.e. take the pseudo-$C_L$ of the masked map. The $f_\text{sky}(L)$ shown in Fig.~\ref{fig:Lfsky} is the average ratio of the output to the input power, as a function of the multipole moment $L$. 
\section{Analysis masks}
\label{ax:masks}
Here we vizualize all the analysis masks we used in this paper: the fiducial temperature and polarization masks, as well as the two test masks used in \S\ref{sec:systematics}.
\begin{figure*}[htbp]
\includegraphics[height=4.5cm,keepaspectratio=true]{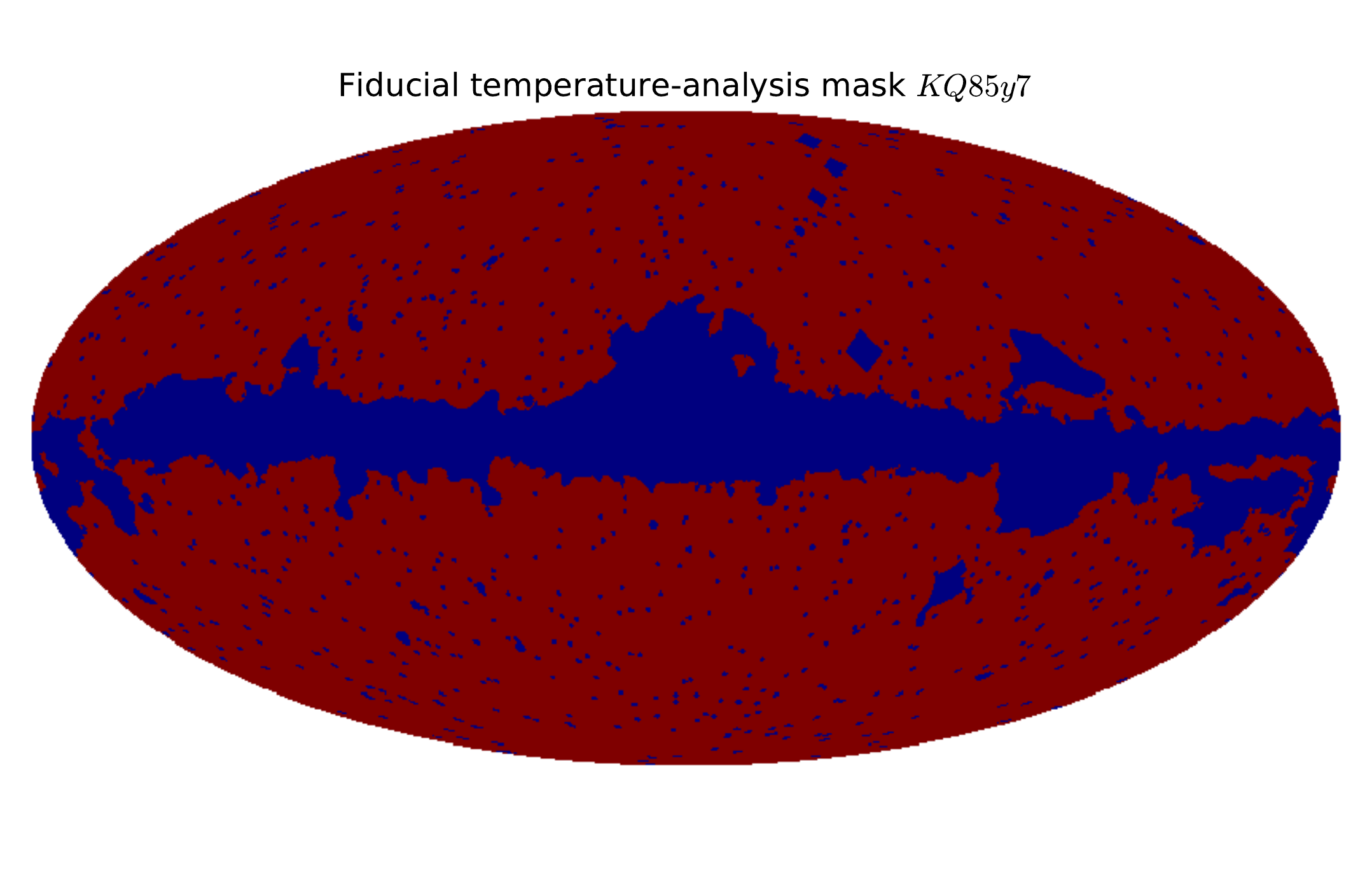}
\includegraphics[height=4.5cm,keepaspectratio=true]{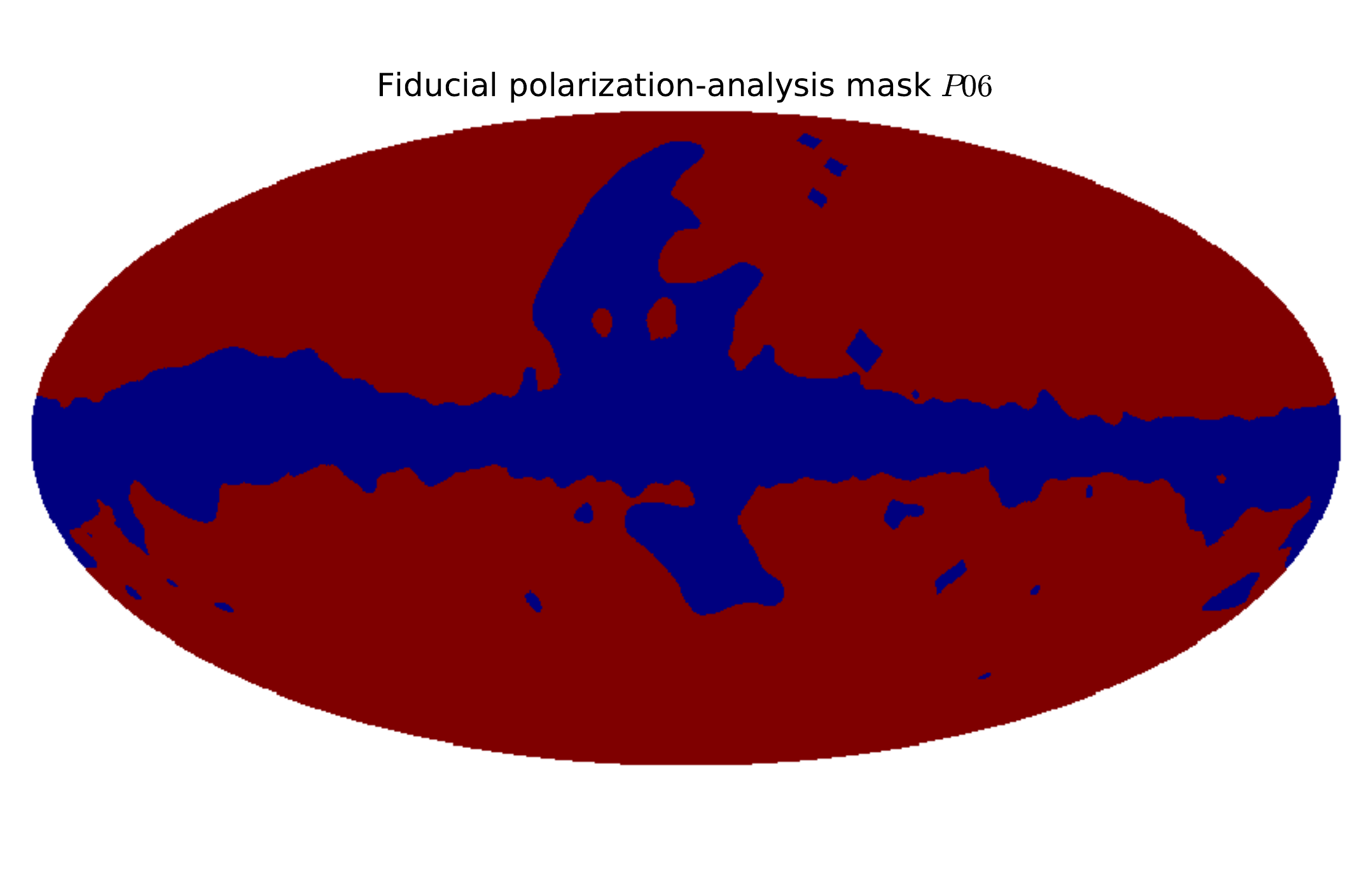}
\includegraphics[height=4.5cm,keepaspectratio=true]{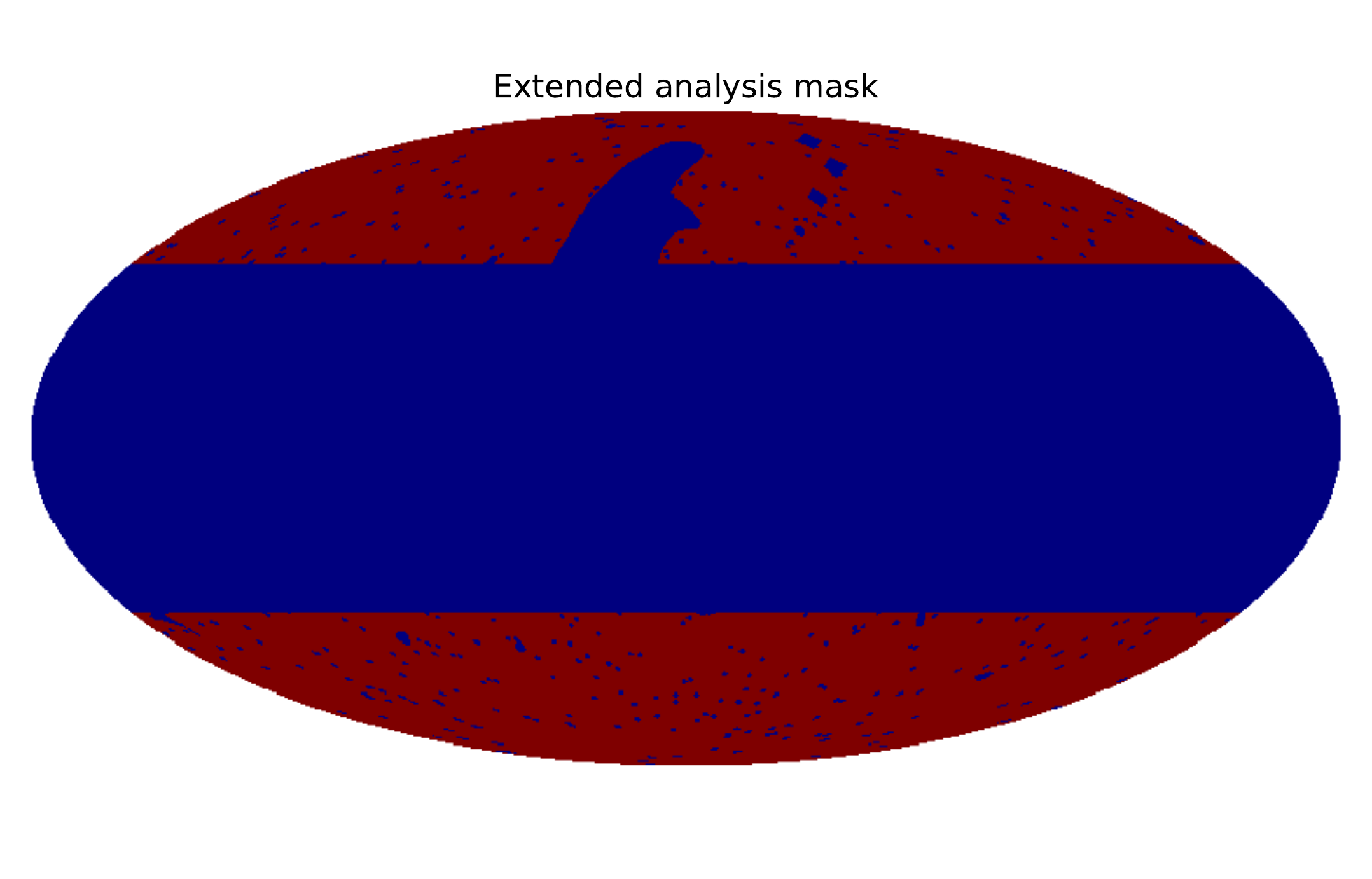}
\includegraphics[height=4.5cm,keepaspectratio=true]{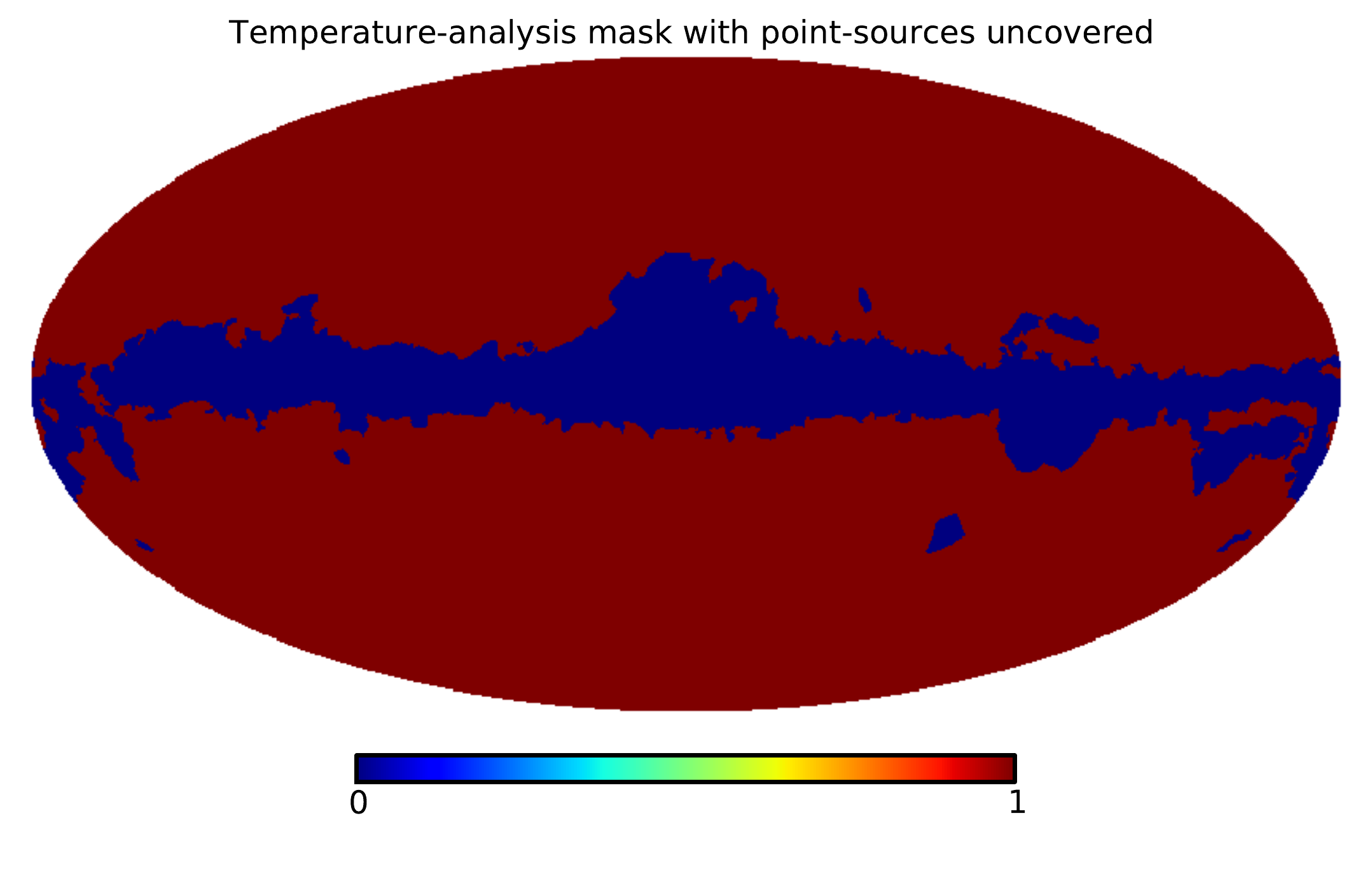}
\caption{Fiducial temperature-analysis mask (with $\sim 78\%$ of the sky admitted), fiducial polarization-analysis mask ($\sim 73\%$), extended mask ($\sim 33\%$), and the temperature-analysis mask with point sources uncovered ($\sim 82\%$). Fraction of the sky admitted of the combined fiducial masks for polarization and temperature is $\sim 68\%$, which is twice the sky admitted as compared to the extended mask. \label{fig:masks}}
\end{figure*}

\end{document}